\documentclass[aps,prd,preprint]{revtex4-2}

\usepackage{amsmath,amssymb}
\usepackage{graphicx}
\usepackage{booktabs}
\usepackage[hidelinks]{hyperref}
\hypersetup{
  pdftitle={Nonuniform Near-Extremal Limits of High-Order Kerr Photon Rings: Critical Inclination, Scaling Laws, and Source Selection},
  pdfauthor={Ren-Qi Pan},
  pdfsubject={Near-extremal Kerr lensing in a finite observer sky},
  pdfkeywords={Kerr black holes, photon rings, gravitational lensing}
}

\graphicspath{{figures/}}
\newcommand{\dif}{\mathrm{d}}
\newcommand{\Mtime}{\mathsf{t}}
\newcommand{\RR}{\mathcal{R}}
\newcommand{\phd}{\Delta\phi}

\begin{document}

\title{Nonuniform Near-Extremal Limits of High-Order Kerr Photon Rings: Critical Inclination, Scaling Laws, and Source Selection}
\author{Ren-Qi Pan}
\email[E-mail: ]{renqipan@zju.edu.cn}
\affiliation{Hangzhou Parity Microelectronics Technology Co.,~Ltd., Hangzhou, Zhejiang 311611, China}
\date{November 30, 2026}

\begin{abstract}
We compute the solid angle of returning Kerr rays of one exact polar order and ask what their travel times reveal about the source. Near extremality, the relevant near-horizon spherical orbits become weakly unstable. Holding $s=n\kappa^{1/3}$ fixed, where $n$ counts polar turns and $\kappa=\sqrt{1-a_{\rm Kerr}^2/M^2}$, we find that a cubic radial potential sets the width of the order. Its solid angle scales as $\kappa^{7/3}$ at inclinations fixed above the critical viewing angle and as $\kappa^2$ inside an $O(\kappa^{2/3})$ layer, where the two screen branches meet. At a moving subcritical visibility edge the area is exponentially small and the rescaled signed screen height is Gaussian. On the time fluctuation scale, the travel time decreases quadratically with absolute screen height. A fixed source mask of positive limiting occupation and boundary of zero source measure preserves the leading normalized time law in both the critical and moving subcritical regimes. A simultaneous flash from a narrowing equatorial strip can retain source longitude harmonics in the folded arrival phase. For a fixed equatorial annulus, phases vary with emission radius, and intervening disk crossings reweight the radii that contribute. Exact-ray quadratures test these predictions. We give outward-rounded enclosures for finite-spin areas and a mixed-precision interval for a cap occupation fraction.
\end{abstract}

\maketitle
\clearpage

\section{Introduction}

The Event Horizon Telescope resolved ring-shaped millimeter emission around a central brightness depression in M87$^*$ and Sagittarius~A$^*$ \cite{EHTM87Shadow,EHTSgrAShadow}. The brightness reflects both light propagation and the emitting plasma \cite{GrallaHolzWald}. In Kerr, directions asymptotic to unstable spherical photon orbits form a critical curve on the sky \cite{BardeenKerrGeodesics,Teo,GrallaLupsascaLensing}. Nearby rays can spend many cycles near a spherical orbit before escaping, producing increasingly narrow subrings. Their shapes and contributions at long interferometric baselines can probe the spacetime \cite{JohnsonEtAl,GrallaLupsascaMarrone,PaugnatEtAl}. We study the solid angle and travel times of one order in this sequence.

At fixed subextremal spin, a ray close to a regular unstable spherical orbit has a logarithmically long dwell time near the radial double root \cite{BozzaStrong,GrallaLupsascaLensing}. The radial growth exponent $\gamma$ per polar half-period sets the factor $e^{-\gamma}$ by which successive subrings narrow locally. The azimuthal and temporal lapses per half-period set their relative rotation and delay \cite{GrallaLupsascaLensing,JohnsonEtAl,HadarHolography}. For general observer inclinations, Yan et al. derived first-order shifts in the critical curve and these photon-ring parameters due to radial deformations of Kerr that preserve separability and the Kerr angular structure \cite{YanLiYue}. Strong-deflection expansions also describe inclined Kerr observers perturbatively in spin and sources at finite distances \cite{BozzaDeLucaScarpetta,BozzaScarpetta}. Near extremality, orbits close to the horizon can become weakly unstable even as $n$ becomes large. When $n\gamma$ stays finite, the return can sample appreciable variation of the radial curvature over the excursion.

For black-hole mass $M$ and spin parameter $a_{\rm Kerr}$, write $\kappa=\sqrt{1-a_{\rm Kerr}^2/M^2}$. The prograde equatorial photon orbit lies a distance of order $M\kappa$ in Boyer--Lindquist radius from $r=M$ \cite{BardeenPressTeukolsky,Hod}. Inclined spherical orbits have a transition at a different radius, of order $M\kappa^{2/3}$ from $r=M$, at a critical orbital inclination defined by the axial angular momentum and Carter constant \cite{Hod}. Igata, Ishihara, and Yasunishi located the geometric boundary between the exterior horizon and cardioid classes at $(r_{\rm cr}^+-M)/M=2^{1/3}\kappa^{2/3}$ \cite{IgataIshiharaYasunishi}. The radial potential at the corresponding extremal impacts has a triple root at the horizon \cite{ChenHsiehLee}. Writing $x=(r_c-M)/M$ for a spherical radius $r_c$, the near-extremal curvature scales as $x+\kappa^2/x^2$. Its two terms compete at $x=O(\kappa^{2/3})$, so the instability per half-period is $O(\kappa^{1/3})$. This gives the order scale $n=O(\kappa^{-1/3})$.

Visibility also depends on the observer's inclination. Near-horizon expansions for distant observers give line spectra and images of orbiting sources \cite{PorfyriadisShiStrominger,LupsascaPorfyriadisShi,GrallaLupsascaStrominger}. The near-horizon extremal Kerr line (NHEKline) has nonzero extent for viewing angle $i_*<i\leq\pi/2$ from the spin axis, where $i_*=\arccos\sqrt{2\sqrt{3}-3}\simeq47.06^\circ$ \cite{GrallaLupsascaStrominger,IgataIshiharaYasunishi}. At the threshold the line shrinks to a point and the two signs of the local polar momentum meet at zero. The ray still has finite polar amplitude along its path. The square-root dependence of screen height on the polar potential at the observer creates a fold in the impact-parameter parametrization of sky area.

For prograde equatorial rays, strong-deflection results give the extremal deflection law \cite{FeleppaBozzaDeGraaf}, the limit at fixed $n_{\rm wind}\kappa$ for azimuthal winding count $n_{\rm wind}$ \cite{Chowdhury,GonzalezMoyaEtAl}, and polar focusing for small deviations from the equatorial plane \cite{GonzalezMoyaEtAl}. We consider finite polar amplitudes and fix $s=n\kappa^{1/3}$ for polar-turn count $n$. The solid angle refers to a single observer's sky. The symplectic photon-shell volume counts trajectories in phase space \cite{ShirHadar}.

Our reference geometry uses a source sphere at $10M$ and a zero-angular-momentum observer (ZAMO) at $30M$. We follow rays that turn outside the horizon in a band adjacent to the spherical-orbit curve in impact space. The sphere gives a fixed radial section at which to record the emission latitude, longitude, and time. Lensing bands for an equatorial disk can be indexed by equator crossings \cite{PaugnatEtAl,CardenasLupsascaZhu}. We count polar turning points over both radial legs. The turns and crossings interleave, so their counts can differ by one, depending on the endpoint phases. For each sign of the observer's polar momentum we integrate over the complete phase interval for exactly $n$ turns, which fixes the width in the sky measure.

At fixed positive $s$ the radial excursion is comparable to the offset of the spherical orbit from $r=M$. The quadratic and cubic terms then enter at the same order and determine the width in impact space. At fixed inclinations above $i_*$ the projection is regular and the area scales as $\kappa^{7/3}$. Inside the $O(\kappa^{2/3})$ inclination layer, the fold contributes an extra factor of $\kappa^{-1/3}$. In the moving subcritical regime, the minimum accessible instability near the horizon is at the inner edge. There a finite-spin change in the instability can give an order-one correction to the exponentially small area. Our contribution is to use the known spherical-orbit geometry to derive the complete one-order width $W_s$, the inclination crossover $C(s,\nu)$, and the relative area law at the moving edge, and to determine the associated geometric time distribution and source response.

Delays between relativistic images, variability correlations across subrings, and light echoes probe strongly lensed photon paths \cite{BozzaManciniDelay,WongGlimmer,HadarAutocorrelations,WongEchoes}. Here we resolve travel times within one order. Rays with spherical anchors farther from the moving edge accumulate less coordinate time per polar traverse. Their screen height scales as the square root of the anchor displacement from the edge, so the time deficit is quadratic in height. A rapidly swept source longitude averages over a full order. A fixed footprint of positive limiting occupation and boundary of zero measure changes the geometric weight while preserving the leading normalized time law. A simultaneous flash from a narrowing equatorial strip can encode longitude when arrival times are folded at the corotation phase.

Relativistic thin-disk transfer functions combine the map from the sky to the equator with local emission and redshift \cite{CunninghamDisk,BeckwithDone}. Using the intersection geometry, we show how an extended emitter changes the folded response. Different emission radii contribute with different phases, so a wider annulus can intercept more rays while reducing the normalized modulation. Intervening disk crossings change which radii reach the observer and therefore change the phase average.

\section{Ray family and the exact polar bin}
\label{sec:setup}

Use $G=c=1$ and measure radii and time in units of $M$. Set $a=\sqrt{1-\kappa^2}>0$, $R_s=10$, $R_o=30$, and take the observer inclination $i\in[\pi/6,\pi/2]$. For positive energy $E$, the impact constants are $\ell=L_z/(ME)$ and $q=Q/(M^2E^2)$. In the dimensionless Mino parameter $\lambda$, the separated null equations are \cite{Carter,Mino,GrallaLupsascaExterior}
\begin{align}
 \left(\frac{\dif r}{\dif\lambda}\right)^2
 &=\RR=P^2-\Delta\bigl[q+(\ell-a)^2\bigr],
 & P&=r^2+a^2-a\ell,\quad \Delta=(r-1)^2-\kappa^2,\\
 \left(\frac{\dif\theta}{\dif\lambda}\right)^2
 &=\Theta=q+a^2\cos^2\theta-\ell^2\cot^2\theta .
 \label{eq:potentials}
\end{align}
Follow an inward auxiliary ray from the observer to its accessible outer turn $r_t>1+\kappa$, then out to $R_s$. Reversing the parameter and applying the Kerr isometry $(t,\phi)\mapsto(-t,-\phi)$ gives the corresponding future-directed ray from source to observer. The radial and polar paths run in reverse order, with the same impact constants. The elapsed Mino parameter $\Lambda$, Boyer--Lindquist time $\Mtime=\Delta t_{\rm BL}/M$, and unwrapped azimuthal lapse $\phd$ are obtained from
\begin{align}
 \Lambda&=\sum_{R=R_s,R_o}\int_{r_t}^R \frac{\dif r}{\sqrt{\RR}},\\
 \frac{\dif\Mtime}{\dif\lambda}&=K_t+A_t,
 & K_t&=\frac{(r^2+a^2)P}{\Delta}, & A_t&=a\ell-a^2\sin^2\theta,
 \label{eq:time-kernel}\\
 \frac{\dif\phi}{\dif\lambda}&=K_\phi+A_\phi,
 & K_\phi&=\frac{aP}{\Delta}, & A_\phi&=\ell\csc^2\theta-a .
 \label{eq:lapses}
\end{align}
For a fixed observer longitude, $\phi_s=\phi_o-\phd\bmod 2\pi$.

The double-root conditions at the spherical anchor $r_c=1+x$ give \cite{Teo}
\begin{equation}
 \ell_c=\frac{2-x^2+\kappa^2+2\kappa^2/x}{a},\qquad
 q_c=\frac{(1+x)^3}{a^2}\left(3-x-\frac{4\kappa^2}{x^2}\right).
 \label{eq:spherical}
\end{equation}
On the exterior arc $q_c\geq0$, the spherical anchor $x$ determines the location along the critical curve in impact space. The label $w$ displaces the impacts along an unnormalized outward normal in $(\ell,\sqrt q)$,
\begin{equation}
 \ell=\ell_c+w\frac{\sqrt{1+x}(2-x)}{2a},\qquad
 q=q_c\left(1+\frac{wx}{2(1+x)^{3/2}}\right)^2,\qquad 0<w\leq0.01 .
 \label{eq:normal-band}
\end{equation}
Keep inward directions with $\Theta_i=\Theta(i)\geq0$ and $r_t<R_s$ accessible from $R_o$. Degenerate endpoints have zero solid angle.

For regular rays, $T=\int_{\theta_-}^{\theta_+}\dif\theta/\sqrt\Theta$ is the polar half-period. In elliptic parameter notation, $T=2K(u_+/u_-)/(a\sqrt{-u_-})$ with $u_-<0<u_+$ the roots of $q-(q+\ell^2-a^2)u-a^2u^2$ in $u=\cos^2\theta$. Use a phase in units of $T$, with equator crossings at integers and the northern turn at $1/2$. On the two launch branches $\sigma=\operatorname{sgn}(\dif\theta/\dif\lambda)$ of the auxiliary ray the initial phase is $\varphi_\sigma=1/2+\sigma d_o$, with $0<d_o\leq1/2$. A bin of exactly $n$ turns, excluding launch and including arrival on that path, satisfies
\begin{equation}
 \mathcal E\equiv\frac{\Lambda}{T}+\varphi_\sigma
 =n+j_\sigma-\tfrac12+\eta,\qquad
 j_\sigma=\lfloor\varphi_\sigma+\tfrac12\rfloor,\qquad 0\leq\eta<1 .
 \label{eq:event}
\end{equation}
Thus $|n-\Lambda/T|<1$. At the northern visibility fold both launch phases tend to $1/2$, while $j_-=0$ and $j_+=1$ persist on the respective one-sided branches.

\begin{figure}[tbp]
 \centering
 \includegraphics[width=\linewidth]{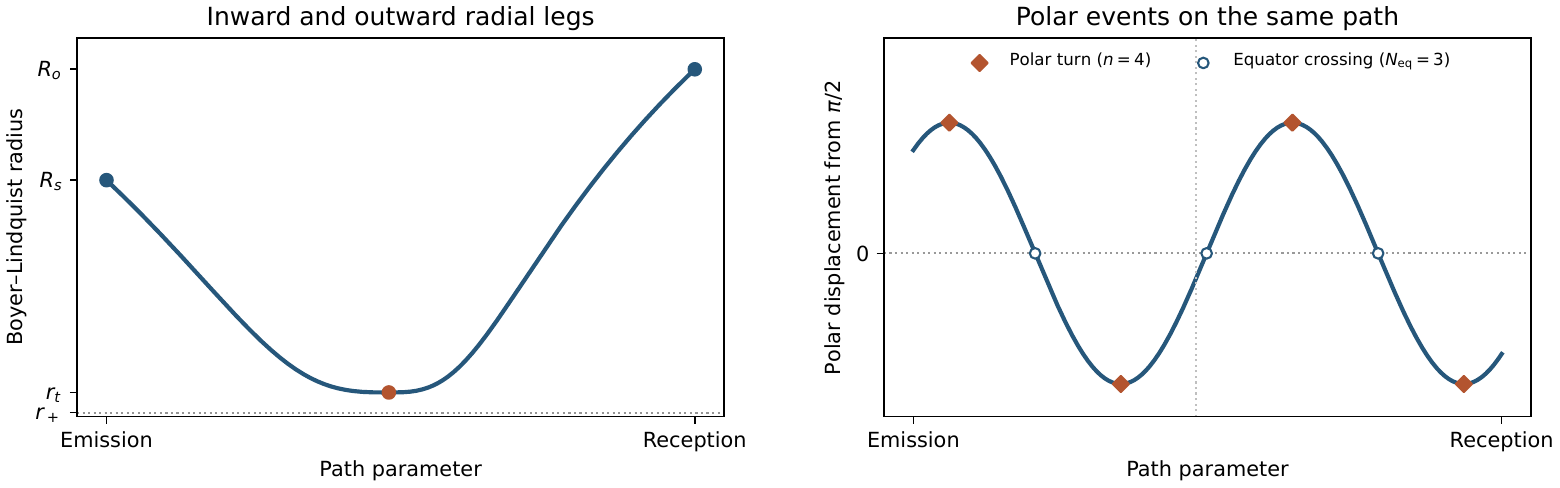}
 \caption{Schematic exterior return in the physical direction. The ray leaves the source sphere inward, turns at $r_t>r_+$, then reaches the observer. The polar motion on the same path illustrates four turns and three equator crossings. The dotted vertical line marks the radial turn in the polar panel.}
 \label{fig:return}
\end{figure}

On the same finite path the equator crossing count is $N_{\rm eq}=\lfloor\mathcal E\rfloor-\lfloor\varphi_\sigma\rfloor$. The turns and crossings interleave, giving $|N_{\rm eq}-n|\leq1$ (Fig.~\ref{fig:return}). Hence the two counts give the same leading scaled order, with $N_{\rm eq}b=nb+O(b)$. The coefficient of one complete bin uses the endpoint phases specified in Eq.~\eqref{eq:event}. Section~\ref{sec:disk} applies this bin convention to an equatorial disk by selecting its emission crossing.

For the finite ZAMO sky \cite{BardeenPressTeukolsky,GrenzebachPerlickLaemmerzahl}, let $\Delta_o=\Delta(R_o)$ and $\mathcal A_o=(R_o^2+a^2)^2-a^2\Delta_o\sin^2 i$. The gnomonic projection gives
\begin{equation}
 C_o=\frac{(R_o^2+a^2\cos^2 i)\sqrt{\Delta_o}}{\sqrt{\mathcal A_o}\sin i},\qquad
 (\alpha,\beta)=\left(-\frac{C_o\ell}{\sqrt{\RR_o}},
             \sigma\sqrt{\frac{\Delta_o\Theta_i}{\RR_o}}\right),
 \label{eq:sky}
\end{equation}
where $\RR_o=\RR(R_o)$. The gnomonic slopes have sky measure $|\dif\alpha\,\dif\beta|/(1+\alpha^2+\beta^2)^{3/2}$. The square root in $\beta$ gives a factor $\Theta_i^{-1/2}$ when converting this area to impact constants. The solid angle for one sign is
\begin{align}
 \dif\Omega_\sigma&=\frac{S_o B_o}{\sqrt{\Theta_i}}\,\dif x\,\dif w,
 & S_o&=\frac{C_o\sqrt{\Delta_o}|D_o-\ell\partial_\ell D_o/2|}
              {2\RR_o^2(1+\alpha^2+\beta^2)^{3/2}},\\
 B_o&=\frac{8(1+x)^{3/2}h}{ax}
 \left(1+\frac{wx}{2(1+x)^{3/2}}\right)
 \left(1+\frac{3wx}{8h\sqrt{1+x}}\right),
 & h&=x+\frac{\kappa^2}{x^2},
 \label{eq:jacobians}
\end{align}
Here $D_o=\RR_o+\Delta_o\Theta_i=P(R_o)^2-\Delta_o(\ell\csc i-a\sin i)^2$. The factor $S_o$ is regular at the fold, and $B_o=|\partial(\ell,q)/\partial(x,w)|$ converts the area in impact constants to the radial-anchor and width labels. Set $v=\log w$. At fixed radial or fold label, $|\dif w|=w\,\dif\eta/|\partial_v\mathcal E|$ converts a phase interval into a width of the order in these labels. Summing both signs and all physical branches defines the one-order solid angle $G_{\kappa,n}(i)$. Write $P_{\kappa,n,i}$ for this sky measure restricted to the order and normalized by $G_{\kappa,n}(i)$.

\section{Cubic return clock and solid-angle crossover}
\label{sec:inclination}

\subsection{Joint radial and order scaling}

For $w=0$ and $y=r-r_c$, the potential is exactly $\RR(r_c+y)=4r_chy^2+4r_cy^3+y^4$. The spherical Mino instability is $\lambda_r=\sqrt{\RR''(r_c)/2}=2\sqrt{r_ch}$, and the accumulated exponent per polar half-period is $\gamma_\kappa(x)=\lambda_r T_c$, with $T_c$ evaluated at the spherical impacts. The terms $x$ and $\kappa^2/x^2$ in $h$ balance on the scale $x\sim\kappa^{2/3}$. Introduce
\begin{equation}
 b=\kappa^{1/3},\qquad s=nb,\qquad X=x/b^2,\qquad Y=(r-1)/b^2 .
\end{equation}
Set $H=X+X^{-2}$ and $g(X)=8X-4X^{-2}$. For finite positive $X$, $\gamma_\kappa/b\to2T_*\sqrt H$, where $T_*=T(1,2,3)$ in $(a,\ell,q)$ notation. The minimum of $H$ at $X_*=2^{1/3}$ gives the weakest instability to leading order. Its leading radius coincides with the geometric division of spherical-orbit families found by Igata et al. \cite{IgataIshiharaYasunishi}. Here the minimum follows from the radial growth exponent, which sets the ordering of nearby returning rays. At an excursion $y=O(b^2)$, the quadratic and cubic terms of the potential are both of order $b^6$. The leading displacement of the potential from Eq.~\eqref{eq:normal-band} is $-4w(r-1)$ on this scale, so an offset $w=b^4\delta$ affects the radial turn at the same order. On positive compact sets, $\ell=2+O(b^4)$, $q=3+b^2g(X)+O(b^4)$, and
\begin{equation}
 b^{-6}\RR(1+b^2Y)=F(Y;X,\delta)+O(b^2),\qquad
 F=4\bigl[(Y-X)^2(Y+X^{-2})-\delta Y\bigr].
 \label{eq:cubic}
\end{equation}
The outer root is $Y_t=X+d$, where $\delta=d^2(d+H)/(d+X)$ is strictly increasing in $d>0$. The limiting scaled clock is
\begin{equation}
 J_0(X,\delta)=\frac{2}{T_*}\int_{Y_t}^\infty\frac{\dif Y}{\sqrt F}
 =\frac{4}{T_*}\int_0^\infty\frac{\dif u}{\sqrt{A+B u^2+4u^4}},
 \label{eq:clock}
\end{equation}
where $Y=Y_t+u^2$, $A=4(3d^2+2Hd-\delta)$, and $B=4H+12d$. Both $A$ and $B$ increase with $d$. Increasing the impact offset moves the turn farther from the spherical orbit and decreases the scaled half-period count from infinity to zero. Its inverse and event weight are
\begin{equation}
 J_0(X,\delta_s(X))=s,\qquad
 W_s(X)=\frac{\delta_s(X)}{-\partial_{\log\delta}J_0(X,\delta_s(X))}
       =-\partial_s\delta_s(X)>0 .
 \label{eq:weight}
\end{equation}
For fixed $X$, the logarithmic saddle law describes the corner $\delta\downarrow0$, where the turn lies close to the double root in the scaled radius. At finite positive $\delta$, the excursion enters the leading integral. The inverse slope $W_s$ measures the width swept out for a given interval in the scaled count $s$.

The exact integrals have an additional $b^2u^6$ term in the squared denominator and upper endpoints of order $b^{-1}$. Their scaled sum and log-width derivative converge on compact trial labels. Equation~\eqref{eq:event} then gives $w/b^4=\delta_s(X)+o(1)$ and $b\,\partial_v\mathcal E=\partial_{\log\delta}J_0(X,\delta_s)+o(1)$ at the roots, with $X$ taken at its limiting value on each chart. The width element is $|\dif w|=b^5[W_s(X)+o(1)]\dif\eta$, since one polar turn spans $b$ in the scaled count. Appendix~\ref{app:bounds} controls the rest of the physical order.

\subsection{Critical layer and ordinary inclinations}

At the limiting impacts, set
\begin{equation}
 \Theta_*(i)=3+\cos^2 i-4\cot^2 i,\qquad
 i_*=\arccos\sqrt{2\sqrt3-3},\qquad c_i=\Theta_*'(i_*)>0 .
\end{equation}
The sky in Eq.~\eqref{eq:sky} admits the limiting impacts $(a,\ell,q)=(1,2,3)$ when $\Theta_*(i)\geq0$. The observer threshold $i_*$ follows from the local polar momentum squared at the observer. An inclination shift of order $b^2$ changes this momentum squared by the same order as the drift of $q_c$ across $X=O(1)$. For $i=i_*+b^2\nu$, the compact-label limit of $\Theta_i/b^2$ is $g(X)+c_i\nu$. Since $g$ increases with $X$, decreasing $i$ raises the minimum visible anchor $L=g^{-1}(-c_i\nu)$. In the regular fold coordinate $Z=\sqrt{\Theta_i}/b$, visible directions have the limiting label $X_0=g^{-1}(Z^2-c_i\nu)$. The one-sign density in $(Z,\eta)$ is
\begin{equation}
 \rho_\sigma=\frac{2bS_oB_o}{|\Theta_x|}\,\frac{w}{|\partial_v\mathcal E|_Z},
 \qquad \Theta_x=\partial_x\Theta_i\big|_w .
 \label{eq:fold-density}
\end{equation}
Indeed $\partial_Z x|_w=2b^2Z/\Theta_x$. Both $B_o$ and $\Theta_x$ tend to $g'(X_0)$ there. The limit of $\rho_\sigma/b^6$ is $2S_*W_s(X_0)$, where $S_*=S_o$ at $(a,\ell,q,i)=(1,2,3,i_*)$. At fixed visible $i>i_*$ the one-sign density in $(X,\eta)$ instead tends after division by $b^7$ to $S_*(i)g'(X)W_s(X)/\sqrt{\Theta_*(i)}$, with $S_*(i)$ evaluated at the same limiting impacts.

For $s$ in compact positive intervals the full exact-order area obeys
\begin{align}
 G_{\kappa,n}(i_*+b^2\nu)&=b^6C(s,\nu)[1+o(1)],&
 C(s,\nu)&=4S_*\int_0^\infty W_s(X_0)\,\dif Z, \label{eq:critical}\\
 G_{\kappa,n}(i)&=b^7C_{\rm ord}(s,i)[1+o(1)],&
 C_{\rm ord}(s,i)&=\frac{2S_*(i)}{\sqrt{\Theta_*(i)}}\,\mathcal M(s),
 \label{eq:ordinary}
\end{align}
Here $\mathcal M(s)=\int_0^\infty g'(X)W_s(X)\,\dif X$. The critical result is uniform for bounded $\nu$, and the ordinary result on compact subsets of $(i_*,\pi/2]$. Both use the actual $s=nb$. At a regular projection, the anchor span $b^2$ and bin width $b^5$ give area $b^7=\kappa^{7/3}$. Near the fold, $\Theta_i=O(b^2)$ enhances the projected area by $b^{-1}$, giving $b^6=\kappa^2$. The density in $Z$ is regular there. Replacing $W_s$ by $\delta_s$ gives the limiting cumulative coefficient, whose negative $s$ derivative yields the model bin coefficient. The exact coarea density and its relative tail estimates establish the limit for the physical bin.

\begin{figure}[tbp]
 \centering
 \includegraphics[width=\linewidth]{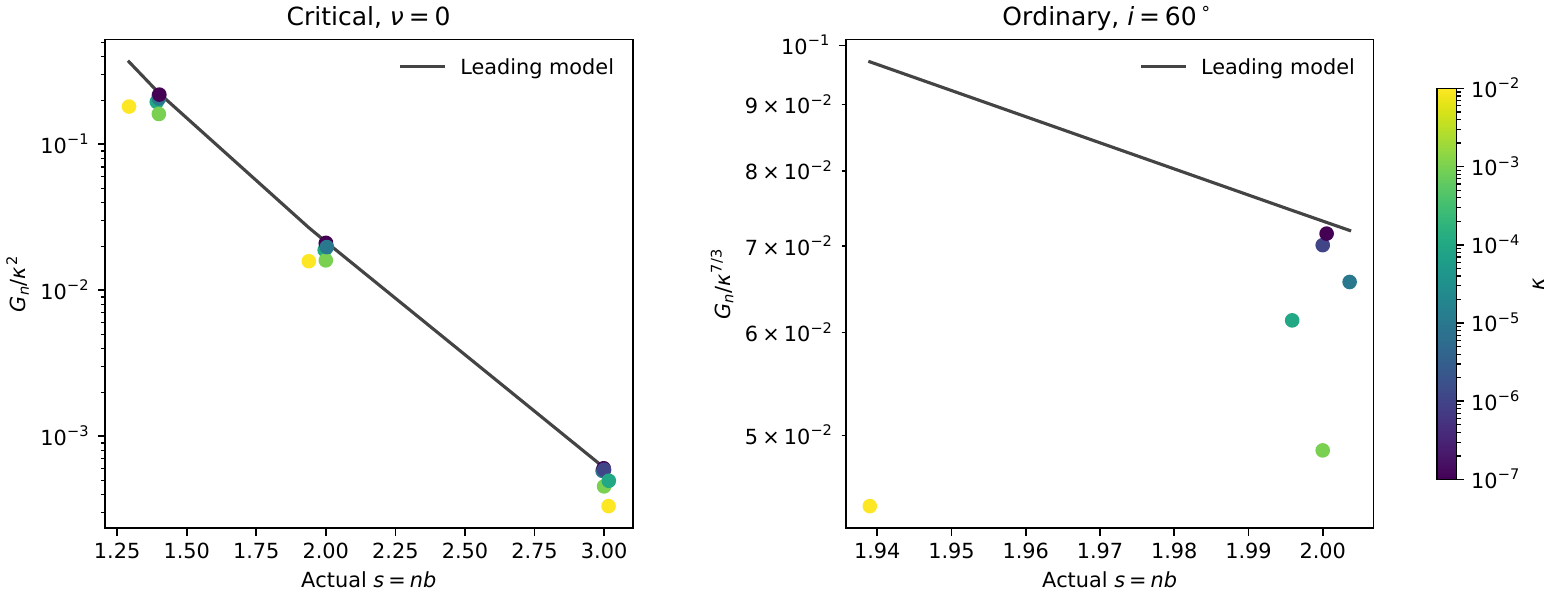}
 \caption{Scaled solid angles for one exact polar order at $i=i_*$ and $i=60^\circ$. The points are exact-geodesic quadratures at the actual $s=nb$. The bars show the differences between the two finest quadratures. The curves are the leading coefficients in Eqs.~\eqref{eq:critical} and \eqref{eq:ordinary}.}
 \label{fig:collapse}
\end{figure}

\subsection{Matching to both sides of the layer}

Changing from $Z$ to $X$ in Eq.~\eqref{eq:critical} gives the smooth positive coefficient
\begin{equation}
 C(s,\nu)=2S_*\int_L^\infty
       \frac{g'(X)W_s(X)}{\sqrt{g(X)-g(L)}}\,\dif X .
 \label{eq:abel}
\end{equation}
In either radial tail, $X\to0$ or $X\to\infty$, the inverse has
\begin{equation}
 \delta_s\sim64\,\frac{H^3}{X}e^{-2sT_*\sqrt H},\qquad
 W_s\sim128T_*\,\frac{H^{7/2}}{X}e^{-2sT_*\sqrt H}.
 \label{eq:modeltails}
\end{equation}
These equivalents follow from the elliptic logarithm (Appendix~\ref{app:clock}). The first inclination asymptotic follows by taking the large positive $\nu$ limit at fixed $X$ and bounding the small-$X$ fold tail. On the other side $L\sim c_i|\nu|/8$. Substitution of $X=(\sqrt L+y)^2$ leaves an endpoint Laplace integral in $y$ with weight $y^{-1/2}e^{-2sT_* y}$. The results are
\begin{align}
 C(s,\nu)&\sim\frac{2S_*\mathcal M(s)}{\sqrt{c_i\nu}},
 &&\nu\to+\infty, \label{eq:pos-model}\\
 C(s,\nu)&\sim1024S_*T_*\sqrt{\frac{\pi}{2sT_*}}\,
 L^{11/4}e^{-2sT_*\sqrt L},
 &&\nu\to-\infty . \label{eq:neg-model}
\end{align}
In the separate large-$s$ model limit, the area peaks when the fold approaches the least unstable anchor $X_*$. The peak tends to $\nu=-g(X_*)/c_i\simeq-0.58624$, and the full width at half maximum in $\nu$ is asymptotic to $3.04558/\sqrt{s}$ (Appendix~\ref{app:clock}). Figure~\ref{fig:crossover} tests the critical coefficient and its subcritical matching with finite-spin rays.

The physical positive overlap is $1\ll\nu\ll\kappa^{-2/3}$. The fold moves to small $X$, where the growing $X^{-2}$ term in $H$ suppresses the area. On fixed $X$ intervals $\Theta_i/\Theta_*(i)\to1$, so the projection approaches the ordinary inclination law. The area is asymptotic to both $b^7C_{\rm ord}(s,i)$ and $b^6C(s,\nu)$ (Appendix~\ref{app:bounds}).

For the moving subcritical limit, take
\begin{equation}
 L\to\infty,\qquad u_e=b^2L\to0,\qquad s\ \text{in a compact positive interval}.
 \label{eq:moving-limit}
\end{equation}
Let $x_e$ be the inner finite-spin spherical visibility edge, determined by $\Theta_i(x_e,0)=0$, and set
\begin{equation}
 \gamma_e=\gamma_\kappa(x_e),\qquad
 \ell_e=(n\gamma_\kappa'(x_e))^{-1}\sim\frac{\sqrt{x_e}}{nT_*}\ll x_e.
 \label{eq:edge-width}
\end{equation}
The visibility edge $x_e\sim u_e$ lies on the increasing branch of the instability. The exponential narrowing selects the scale $\ell_e$ over which $n\gamma_\kappa(x)$ changes by order unity. Define the spherical label $x_0$ by $\Theta_i(x,w)=\Theta_i(x_0,0)$. Uniform logarithmic matching of the two-leg clock gives, on $x_0=x_e+O(\ell_e)$,
\begin{equation}
 w=64x_0^2 e^{-n\gamma_\kappa(x_0)}[1+o(1)],\qquad
 |\partial_v\mathcal E|_{\Theta_i}=\gamma_\kappa(x_0)^{-1}[1+o(1)] .
 \label{eq:edge-width-law}
\end{equation}
The sky factor is asymptotic to $S_*\sqrt{8/(x_0-x_e)}$. On putting $z_e=(x_0-x_e)/\ell_e$, the width density contributes $64x_e^2\gamma_e e^{-n\gamma_e}e^{-z_e}$. Integrating the square-root factor and summing the signs gives the relative area law
\begin{equation}
 G_{\kappa,n}(i)\sim512 S_*\sqrt{2\pi T_*}\,
 x_e^{11/4}n^{-1/2}e^{-n\gamma_e}.
 \label{eq:edge-area}
\end{equation}
The fold and instability factors give normalized density $z_e^{-1/2}e^{-z_e}/\sqrt\pi$, or $\mathrm{Gamma}(1/2,1)$ with shape $1/2$ and rate one. In the limit, $z_e$, the equiprobable screen sign, and the uniform bin phase are independent. Appendix~\ref{app:edge} controls the tails relative to the exponentially small area. On the concentration layer $x-x_0=O(w)=o(\ell_e)$, so $x$ gives the same radial limit.

Although $\gamma_e$ is small, the factor $n$ can amplify a finite-spin correction in the exponent. Expanding the visibility equation and the spherical period gives
\begin{align}
 x_e&=u_e+K_e u_e^2+O(u_e^3+u_e^2 L^{-3}), & K_e&=\frac{3-\sqrt3}{12},\\
 \frac{\gamma_e}{2T_*\sqrt{u_e}}
 &=1-\frac{1+\sqrt3}{24}u_e+O(u_e^2+L^{-3}).
 \label{eq:edge-expanded}
\end{align}
With $\tau_{\rm edge}=b^2L^{3/2}$, the resulting matching formula is
\begin{equation}
 \log\frac{G_{\kappa,n}(i)}{b^6 C(s,\nu)}
 =\frac{sT_*(1+\sqrt3)}{12}\,\tau_{\rm edge}+o(1+\tau_{\rm edge}) .
 \label{eq:neg-overlap}
\end{equation}
The parameter $\tau_{\rm edge}$ measures the accumulated first correction to the exponent. The uncorrected critical coefficient gives a relative equivalent in $1\ll-\nu\ll\kappa^{-4/9}$. A finite positive limit of $\tau_{\rm edge}$ gives an order-one multiplicative correction, and Eq.~\eqref{eq:edge-area} holds throughout Eq.~\eqref{eq:moving-limit}. For fixed $i\in[\pi/6,i_*)$, the visible spherical arc stays away from the horizon. Its least instability sets the logarithmic area rate $n^{-1}\log G_{\kappa,n}(i)\to-\Gamma_i$, with $\Gamma_i=\min_{x\in I_i}2\sqrt{x(1+x)}\,T_c(1,x)>0$ on the visible arc $I_i$ at $a=1$ (Appendix~\ref{app:edge}). Table~\ref{tab:limits} gives the domains of the area results.

\begin{table}[tbp]
 \caption{Limits for one-order area. The physical limits take $b\to0$ with $s=nb$ in a compact positive interval.}
 \label{tab:limits}
 \begin{ruledtabular}
 \begin{tabular}{p{.34\linewidth}p{.59\linewidth}}
 Inclination regime & Area law and comparison \\
 \midrule
 Fixed $i>i_*$ & $b^7C_{\rm ord}$ on compact subsets of $(i_*,\pi/2]$. Positive overlap at $1\ll\nu\ll b^{-2}$.\\
 Bounded $\nu=(i-i_*)/b^2$ & $b^6C$. The joint source and lapse law below is stated for $1\leq s\leq6$, $|\nu|\leq1$.\\
 Moving edge, $L\to\infty$, $b^2L\to0$ & Relative law~\eqref{eq:edge-area} with exact finite-spin exponent. Uncorrected critical overlap at $1\ll-\nu\ll b^{-4/3}$.\\
 Fixed $i<i_*$ & $n^{-1}\log G_{\kappa,n}\to-\Gamma_i$ for $i\in[\pi/6,i_*)$.\\
 Large-$s$ model peak & Take $s\to\infty$ in the limiting $C(s,\nu)$ after the critical-layer limit.\\
 \end{tabular}
 \end{ruledtabular}
\end{table}

\section{Critical-layer lapses and source selection}
\label{sec:critical-sources}

At $r=1+b^2Y$, the scaled kernels $b^2K_\phi$ and $b^2K_t$ tend to $2/Y$ and $4/Y$. Their ratio approaches the extremal horizon angular velocity $M\omega_H=1/2$, reflecting the strong dragging of the rays in the near-horizon region. The angular terms contribute $O(b^{-1})$ to the integrated lapses on compact labels. Thus $\kappa\phd$ and $\kappa\Mtime$ tend to $\Psi_0$ and $2\Psi_0$, respectively, evaluated at $(X,\delta_s(X))$, where
\begin{equation}
 \Psi_0(X,\delta)=4\int_{Y_t}^\infty\frac{\dif Y}{Y\sqrt F}.
 \label{eq:psi}
\end{equation}
Freezing the radial kernel at the spherical value replaces $Y^{-1}$ by $X^{-1}$ and overestimates the leading lapse, since $Y\geq Y_t>X$. The residual azimuth in a frame rotating at the extremal horizon angular velocity is $\phd-\Mtime/2=-D/2$. The leading terms cancel in $D$, giving a smaller scale,
\begin{equation}
 D=nH_*+O(1),\qquad H_*=4T_*+\int_{\rm half}(A_t^*-2A_\phi^*)\,\dif\lambda
     =2.1621794914\ldots ,
 \label{eq:gap}
\end{equation}
with the bounded remainder on compact fold cores. The star denotes the limiting impacts and $a=1$. The radial difference kernel is $(r^2+a^2-2a)P/\Delta=4+O(b^2(1+Y))$ on the legs. Its nonconstant part integrates to $O(1)$ with fixed endpoints, while the angular kernel contributes its half-period integral times $n$ plus $O(1)$.

\begin{figure}[tbp]
 \centering
 \includegraphics[width=\linewidth]{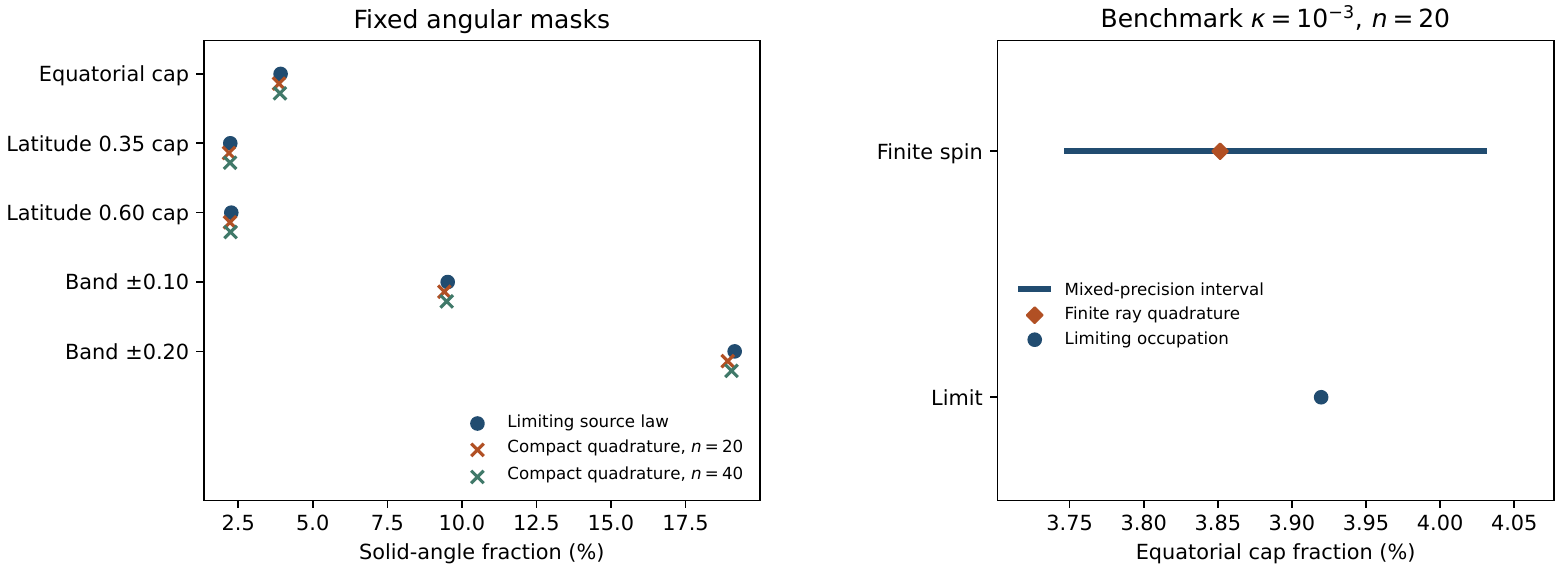}
 \caption{Geometric source selection. The cap centers and radii $(\text{latitude},\text{longitude},\text{radius})$ are $(0,0,0.40)$, $(0.35,0.60,0.28)$, and $(0.60,-1.00,0.22)$ rad. Bands are centered on the equator. The left panel shows the whole limiting law and fractions in compact ray windows at $s=2$, $\nu=0$ with $X_0\leq14$ for $n=20$ and $X_0\leq12$ for $n=40$. The right panel compares the limiting occupation, the finite ray quadrature over the visible interval, and the full-band mixed-precision interval for the equatorial cap.}
 \label{fig:source}
\end{figure}

For $s\in[1,6]$ and $|\nu|\leq1$, the joint leading law $\mu_{s,\nu}$ of the scaled winding and coordinate time $(U,V)=(\kappa\phd/(2\pi),\kappa\Mtime)$ is the image of $4S_*W_s(X_0)\dif Z/C(s,\nu)$ under $Z\mapsto(\Psi_0/(2\pi),2\Psi_0)$, with $\delta=\delta_s(X_0)$. Corotation gives its support on $V=4\pi U$. The emission position has limiting probability measure
\begin{equation}
 \Theta_0(\theta)=3+\cos^2\theta-4\cot^2\theta,\qquad
 p(\dif\theta\,\dif\phi)
 =\frac{\mathbf 1_{i_*<\theta<\pi-i_*}}{T_*\sqrt{\Theta_0(\theta)}}\,
    \dif\theta\,\frac{\dif\phi}{2\pi}.
 \label{eq:source-law}
\end{equation}
The bin samples one polar half-period uniformly in Mino phase, giving integrable latitude enhancements at the polar turns. At fixed fold label the longitude sweeps through order $b^{-2}$ radians, with positive leading slope for $\phd$ from the width derivatives of $J_0$ and $\Psi_0$. The leading scaled lapses stay constant across the bin. This separation and the relative sky tails (Appendix~\ref{app:timing}) yield the product limit
\begin{equation}
 \int f(\theta_s,\phi_s)F_1(U,V)\,\dif P_{\kappa,n,i}
 =p[f]\int F_1\,\dif\mu_{s,\nu}+o(1)
 \label{eq:critical-product}
\end{equation}
for fixed continuous source tests and bounded continuous lapse tests, uniformly on the stated family. For a fixed mask $\mathcal D$ with $p(\partial\mathcal D)=0$ and $p(\mathcal D)>0$, its leading area is $p(\mathcal D)G_{\kappa,n}$ and its normalized lapse law is still $\mu_{s,\nu}$. A cap of radius $\rho_s$ selects $\arccos(\hat{\mathbf n}_s\cdot\hat{\mathbf n}_{\rm cap})\leq\rho_s$ on the unit sphere of source angles. For the equatorial cap of radius $0.4$ rad, Eq.~\eqref{eq:source-law} gives $p(\mathcal D)\simeq0.0391970$ (Fig.~\ref{fig:source}).

\section{Subcritical screen, arrival time, and thin flashes}
\label{sec:subcritical-sources}

\subsection{A Gaussian screen with a quadratic time deficit}

In the moving regime the spherical time and azimuth per polar half-period follow from Eqs.~\eqref{eq:time-kernel} and \eqref{eq:lapses}. For the anchor at $r_c=1+x$ define
\begin{equation}
 f_j(x)=T_c K_j^c(r_c)+
   \int_{\theta_-}^{\theta_+}\frac{A_j^c(\theta)}{\sqrt{\Theta^c(\theta)}}\,\dif\theta,
 \qquad (f_t,f_\phi)=(\tau_\kappa,\varphi_\kappa).
 \label{eq:spherical-lapses}
\end{equation}
The superscript $c$ evaluates the kernels and polar potential at $(\ell_c,q_c)$. Define
\begin{align}
 B_e^2&=\frac{\Delta_o}{\RR_o(x_e,0)}\,\Theta_x(x_e,0)\ell_e,
 & Y_\beta&=\beta/B_e,\\
 D_t&=x_e^{3/2}[n\tau_\kappa(x_e)-\Mtime],
 & D_\phi&=x_e^{3/2}[n\varphi_\kappa(x_e)-\phd].
 \label{eq:deficits}
\end{align}
Here $B_e$ has order $x_e^{1/4}n^{-1/2}$. The reference lapses in Eq.~\eqref{eq:deficits} are those for $n$ polar half-periods on the spherical edge orbit. For a simultaneous flash, a larger $D_t$ means earlier reception. Under the full conditional sky measure,
\begin{equation}
 (Y_\beta,D_t,D_\phi)\ \Longrightarrow\
 \bigl(Y_\infty,\,8+4Y_\infty^2,\,4+2Y_\infty^2\bigr),
 \qquad Y_\infty\sim N(0,1/2).
 \label{eq:screen-time}
\end{equation}
The screen projects $z_e$ to the two signs of $\sqrt{z_e}$. The Jacobian of this change cancels the factor $z_e^{-1/2}$ in the radial density, leaving a Gaussian in the signed height. With $(k_t,k_\phi)=(4,2)$, the spherical derivatives satisfy $x^2 f_j'(x)\to-k_jT_*$. An anchor farther from the edge accumulates less time and azimuth per half-period, yielding the term $k_jz_e$ in the deficit. To compute the constant, subtract the spherical radial kernel at $r_c$ and put $\xi=(r-r_c)/x$. The scaled kernel difference tends to $-k_j\xi/(1+\xi)$ and the scaled potential $x^{-3}\RR$ tends to $4\xi^2(1+\xi)$. Each radial leg extends to larger radii and contributes $-k_j$ to the scaled lapse relative to that at its own spherical label (Appendix~\ref{app:timing}). The two legs therefore add $2k_j$ to the deficit.

The induced time law has $(D_t-8)/4\Rightarrow\mathrm{Gamma}(1/2,1)$. The latest limiting arrivals are near $\beta=0$. Within one order, fixed quantiles of arrival time for a simultaneous flash differ on the scale $Mx_e^{-3/2}$ about a center of order $Mn/x_e$. The expansion $f_j(x_e)=k_jT_*/u_e+d_j+O(u_e+L^{-3})$ gives $d_t\simeq7.45684$ and $d_\phi\simeq2.64733$. Since $n x_e^{3/2}\sim s\tau_{\rm edge}$, the $nd_j$ term can shift the center by the fluctuation scale when $\tau_{\rm edge}$ tends to a positive constant.

For comparison at finite spin put $r_e=1+x_e$, $\lambda_{r,e}=\lambda_r(x_e)$, and use a superscript $e$ for kernels at the edge impacts. The finite spherical intercept and slope in $D_t\simeq I_e+B_tz_e$ are
\begin{align}
 I_e&=x_e^{3/2}\sum_{R_j=R_s,R_o}\int_0^{R_j-r_e}
  \frac{K_t^e(r_e)-K_t^e(r_e+y)}
       {y\sqrt{\lambda_{r,e}^2+4r_e y+y^2}}\,\dif y,\\
 B_t&=-x_e^{3/2}\frac{\tau_\kappa'(x_e)}{\gamma_\kappa'(x_e)} .
 \label{eq:finite-timing}
\end{align}
The intercept integrates the spherical residual on both legs. The slope linearizes the spherical part across the layer $\ell_e$. Both are determined by the scene parameters before integrating the ray distribution.

\begin{figure}[tbp]
 \centering
 \includegraphics[width=\linewidth]{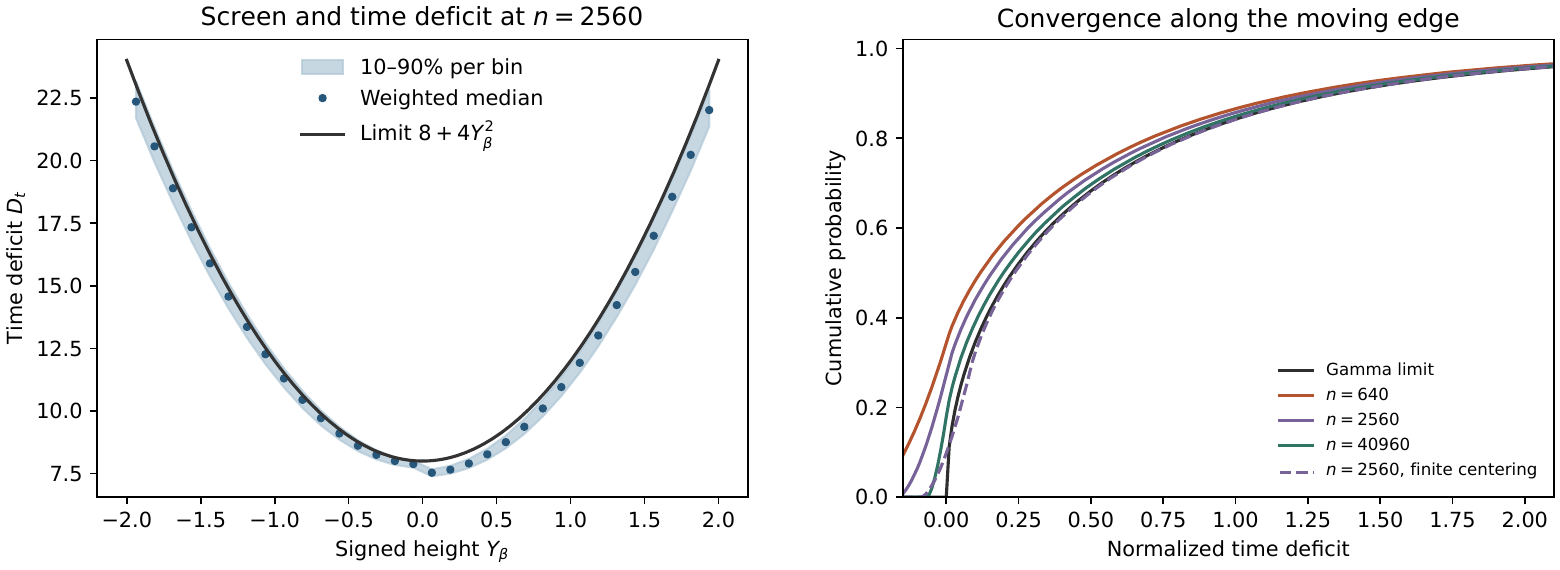}
 \caption{Screen and timing along $s=2$, $\tau_{\rm edge}=0.45$. The left panel shows the weighted median and 10th--90th percentiles in screen bins and the limiting parabola. The right panel compares ray CDFs of $(D_t-8)/4$ with the Gamma CDF. The dashed comparison at $n=2560$ uses $(D_t-I_e)/B_t$ with $I_e=7.68102$ and $B_t=4.00991$.}
 \label{fig:screen-time}
\end{figure}

\subsection{Fixed footprints and a surviving temporal harmonic}

Let the source flash at Boyer--Lindquist time zero. The reception map uses $\Omega_o=M\omega_o=2aR_o/\mathcal A_o$ for the circular ZAMO or $\Omega_o=0$ to hold the reception longitude fixed on the ZAMO reference sky. Both maps use the ZAMO solid angle. An observer with a different four-velocity would evaluate the solid angle in its own tetrad. The limit $\Omega_*=\lim\Omega_o$ satisfies $1-2\Omega_*>0$. The source longitude is $\phi_s=\phi_{o,0}-\Psi\bmod 2\pi$, where $\phi_{o,0}$ is the reception longitude at zero time and $\Psi=\phd-\Omega_o\Mtime$. At fixed fold label,
\begin{equation}
 x_e\partial_\eta\Mtime\to4T_*,\qquad
 x_e\partial_\eta\Psi\to2T_*(1-2\Omega_*) .
 \label{eq:bin-speed}
\end{equation}
The near-turn part of the width derivative gives these slopes (Appendix~\ref{app:timing}). Each bin sweeps through many longitudes at fixed fold label while its leading screen height and time deficit stay fixed. The limiting joint law is thus the product of $p$ in Eq.~\eqref{eq:source-law} and Eq.~\eqref{eq:screen-time}. A fixed mask of positive occupation with $p$-null boundary scales the area by its occupation, preserving the normalized screen and time law. Successive samples of the same source longitude at fixed label and sign are separated by $4\pi M/(1-2\Omega_*)+o(M)$ in coordinate time.

To retain longitude information within the bin, restrict the latitude to $|\theta_s-\pi/2|\le c x_e/2$ with $c>0$ fixed. This samples $O(x_e)$ of the bin phase and an order-one longitude sweep at fixed fold label. The response assigns the weight $f(\phi_s)\dif\Omega$ of each selected ray to its travel time. The profile $f$ is bounded and nonnegative with mean $\bar f=(2\pi)^{-1}\int_0^{2\pi}f\,\dif\phi>0$. We allow continuous profiles and piecewise continuous windows with a longitude-null discontinuity set. Define $H_\kappa=\tau_\kappa-2\varphi_\kappa$, the gap center $D_{\rm c}$, and the circular phase
\begin{align}
 J_*&=\sum_{R=R_s,R_o}\int_1^R\sqrt{\frac{r+3}{r-1}}\,\dif r,
 & D_{\rm c}&=nH_\kappa(x_e)+J_*, \label{eq:gap-center}\\
 \chi&=(1/2-\Omega_o)\Mtime-\phi_{o,0}-D_{\rm c}/2\pmod{2\pi}.
 \label{eq:phase}
\end{align}
The rate $1/2-\Omega_o$ is the extremal horizon angular speed relative to the receiver in units of $M^{-1}$. The strip weighted by $f$ has area $G_{\kappa,n}c x_e\bar f/(T_*\sqrt 3)[1+o(1)]$. Normalizing by this area, for every fixed integer $m$,
\begin{equation}
 \mathbb E_f[e^{im\chi}]\longrightarrow
 \frac{\int_0^{2\pi}f(\phi)e^{-im\phi}\,\dif\phi}{\int_0^{2\pi}f(\phi)\,\dif\phi}
 \cos\!\left(\frac{mH_*}{4}\right).
 \label{eq:harmonics}
\end{equation}
The limiting circular phase is independent of the slow radial variable $z_e$.

To see the two factors, use $\chi=-\phi_s+(D-D_{\rm c})/2\bmod 2\pi$. In the strip, the initial phase tends to $1/2$ and the terminal phase to $n+j_\sigma$. After subtraction of the spherical radial gap, the two-leg contribution tends to $J_*$. On compact concentration intervals with $0<z_e<\infty$, the total gap satisfies
\begin{equation}
 D=D_{\rm c}+(j_\sigma-\tfrac12)H_*+o(1) .
 \label{eq:two-gap}
\end{equation}
At fixed strip latitude on compact interiors of the radial layer, $\partial_{z_e}\Psi\sim -2(1-2\Omega_*)x_e^{-3/2}$. The sweep samples the longitude profile over many periods, while the centered gap loses its $z_e$ dependence. The phase $\chi$ therefore maps the longitude profile to the arrival phase, with equal-weight contributions shifted by $\pm H_*/4$ giving the cosine. The change from bin phase to latitude bounds the strip tails by $O(x_e)$ times the near-layer area tails. The far area is negligible relative to the selected area (Appendix~\ref{app:timing}), giving Eq.~\eqref{eq:harmonics} over the full order.

For a rectangular window of half-width $0.30$ rad the limiting first-harmonic magnitude is $[\sin(0.30)/0.30]\cos(H_*/4)\simeq0.84462$. A window centered at $\phi_{\rm c}$ gives the factor $e^{-im\phi_{\rm c}}$. Uniform longitude gives zero for $m\ne0$. The phase fold for a single flash identifies arrival times separated by $4\pi M/(1-2\Omega_o)$ in Boyer--Lindquist time. Along the circular receiver worldline the corresponding proper-time interval is $4\pi M/[(1-2\Omega_o)u_o^t]$, with $u_o^t=\dif t_{\rm BL}/\dif\tau_o$. The unnormalized harmonic carries the factor $O(x_eG_{\kappa,n})$, so a finite normalized coefficient accompanies an exponentially small geometric response in this regime.

\begin{figure}[tbp]
 \centering
 \includegraphics[width=\linewidth]{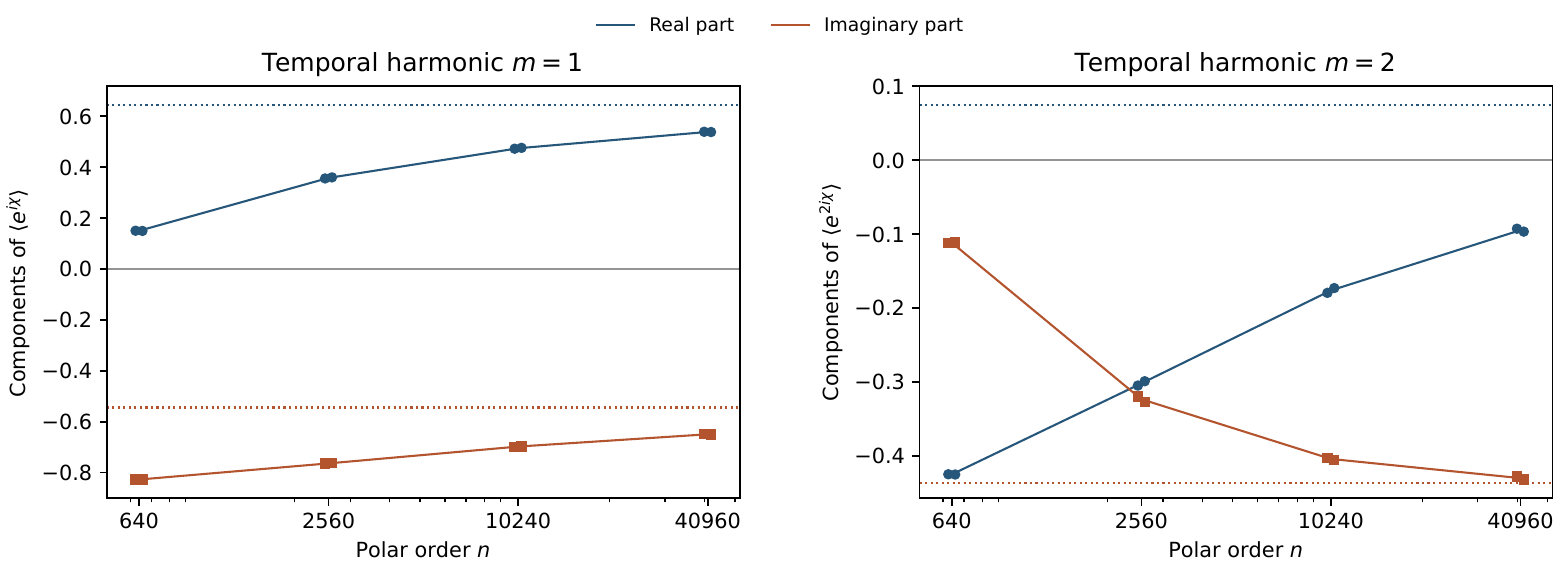}
 \caption{Real and imaginary parts of the normalized arrival-phase coefficients $\mathbb E_f[e^{im\chi}]$ for a strip of latitude width $x_e$ at $s=2$, $\tau_{\rm edge}=0.45$. The circular ZAMO starts at longitude zero, and the source window has center $0.70$ rad and half-width $0.30$ rad. Both panels use $\chi$ from Eq.~\eqref{eq:phase} with the center in Eq.~\eqref{eq:gap-center}. Symbols are two independent scrambles with $2^{17}$ points per sign in the central stratum and separate end strata, covering $0\leq z_e\leq36$. Solid lines connect the two-scramble means. Dotted lines give the limits in Eq.~\eqref{eq:harmonics}. The unnormalized strip area scales as $x_eG_{\kappa,n}$.}
 \label{fig:coherence}
\end{figure}

\subsection{Equatorial disks and occultation}
\label{sec:disk}

Consider a razor-thin equatorial annulus $R_-\le R\le R_+<R_o$, with $R_->1$ fixed as $\kappa\to0$, and use the same band of exterior-return directions. Give one flash the geometric weight $W(R)f(\phi_s)$, with $W$ bounded, nonnegative, and piecewise continuous. Let $\mathcal T\in[0,1]$ be a constant transmission per intervening disk crossing. In backward tracing, a candidate emission intersection has weight $Wf\,\mathcal T^{N_{\rm pre}}$, where $N_{\rm pre}$ counts earlier intersections with the annulus. These are the passages between the candidate emitter and the observer. For $\mathcal T=0$, only the first intersection can contribute. The disk model uses the solid-angle response and a prescribed transmission.

On the two radial legs define the polar phases
\begin{equation}
\begin{aligned}
 \mathcal E_{\rm in}(R)&=\varphi_\sigma+
  \frac{1}{T}\int_R^{R_o}\frac{\dif r}{\sqrt{\RR}},\\
 \mathcal E_{\rm ret}(R)&=\varphi_\sigma+
  \frac{1}{T}\left(\int_{r_t}^{R_o}+\int_{r_t}^{R}\right)
      \frac{\dif r}{\sqrt{\RR}} .
\end{aligned}
 \label{eq:disk-phases}
\end{equation}
Each integer phase reached in the accessible part of the annulus gives one intersection on its leg. Both phases increase along the backward ray, so the intersection radius on either leg is obtained by a scalar monotone inversion. The $n$-turn bin contains the single integer $n+j_\sigma$. For a returning intersection at $R$, count the integers in the inbound phase interval of the annulus. Add the integers from the inner accessible edge on the returning leg up to the emission phase, excluding the emission event. At an inbound emission, only the inbound segment preceding it enters the count. An intersection at the radial turning point is shared by the two legs and contributes once.

At the moving edge the turn lies between the horizon and $R_-$ on each compact concentration interval. Write
\begin{equation}
\begin{aligned}
 \RR_*(R)&=(R-1)^3(R+3),&
 \mathcal Y(R)&=\tfrac12\sqrt{\frac{R+3}{R-1}},\\
 J(R)&=\sqrt{(R-1)(R+3)}+
                4\operatorname{asinh}\left(\frac{\sqrt{R-1}}{2}\right).
\end{aligned}
 \label{eq:disk-kernels}
\end{equation}
The limiting inbound phase is $1/2+[\mathcal Y(R)-\mathcal Y(R_o)]/T_*$. Let $N_{\rm in}^*$ be the number of integers between its values at $R_+$ and $R_-$. Take annular edges for which neither endpoint phase is an integer. Away from return-crossing ties the limiting total number of blockers is
\begin{equation}
 N_{\rm pre}^*(R)=N_{\rm in}^*
  +\left\lfloor\frac{\mathcal Y(R_-)-\mathcal Y(R)}{T_*}\right\rfloor .
 \label{eq:disk-blockers}
\end{equation}
The inbound phase remains bounded on the fixed annulus, so high-order emission intersections in the concentrating layer are on the returning leg.

Changing the emission radius changes the radial Mino lapse by $\dif R/\sqrt{\RR_*(R)}$ at leading order. At a fixed integer crossing the ray width adjusts to compensate. Relative to the reference sphere area, this gives the radial measure $\dif R/(T_*\sqrt{\RR_*(R)})$ before the emission and transmission weights. Keep the phase center $D_{\rm c}$ of Eq.~\eqref{eq:gap-center} with $R_s=10$. The gap at disk emission radius $R$ is $D_{\rm c}+J(R)-J(10)+(j_\sigma-\tfrac12)H_*+o(1)$. After averaging source longitude on the radial concentration scale, the weighted disk area $A_{\rm d}$ and its normalized harmonic $c_m^{\rm d}=\mathbb E_{\rm d}[e^{im\chi}]$ therefore satisfy
\begin{equation}
\begin{aligned}
 \frac{A_{\rm d}}{G_{\kappa,n}}&\longrightarrow
   \frac{\bar f}{T_*}\int_{R_-}^{R_+}\dif\mu_{\rm d}(R),\\
 c_m^{\rm d}&\longrightarrow
  \frac{\widehat f_m}{\bar f}\cos\left(\frac{mH_*}{4}\right)
  \frac{\displaystyle\int_{R_-}^{R_+}
           e^{im[J(R)-J(10)]/2}\,\dif\mu_{\rm d}(R)}
       {\displaystyle\int_{R_-}^{R_+}\dif\mu_{\rm d}(R)},\\
 \dif\mu_{\rm d}(R)&=
  W(R)\mathcal T^{N_{\rm pre}^*(R)}\,\frac{\dif R}{\sqrt{\RR_*(R)}},
 &\widehat f_m&=\frac{1}{2\pi}\int_0^{2\pi}f(\phi)e^{-im\phi}\,\dif\phi .
\end{aligned}
 \label{eq:disk-harmonics}
\end{equation}
Here $0^0=1$, and the normalized law is stated for a positive limiting area ratio. A fixed transparent annulus of positive weight has area of order $G_{\kappa,n}$. Its radial factor averages phases with slope $mJ'(R)/2$, where $J'(R)=\sqrt{(R+3)/(R-1)}$. Phase variation across the annulus can reduce the harmonic magnitude. For annuli narrowing about one radius after the moving-edge limit, the radial factor tends to a phase of unit modulus if the surviving radial weight is positive. A partially transmitting annulus changes the distribution of contributing radii and can change both phase and magnitude. Appendix~\ref{app:disk} derives the radial Jacobian and the averaging limit.

\section{Finite-spin comparisons}
\label{sec:numerics}

An independent ray quadrature brackets the accessible outer root, regularizes the radial legs with $r=r_t+u^2$, and resolves both phase boundaries of each detected width interval separately for each sign. A sine-squared interpolation between visibility edges regularizes the sky integral. The relative refinement difference $\epsilon_Q=|G_{\rm fine}-G_{\rm coarse}|/G_{\rm fine}$ is below $0.005$ over 36 scenes. Figure~\ref{fig:collapse} shows the scaled areas with these per-scene differences. Table~\ref{tab:areas} compares resolution sensitivity with finite-spin prediction errors.

Finite radial endpoints shift the leading area at first order in $b$. For the critical range $1\le s\le6$, $|\nu|\le1$ and the ordinary compact family, let $(m,\mathcal C)=(6,C)$ and $(7,C_{\rm ord})$, respectively. Then
\begin{equation}
 \begin{aligned}
 b^{-m}G_{\kappa,n}&=\mathcal C(s+bB_R)+O\bigl(b^2(1+|\log b|)^N\bigr),\\
 B_R&=\frac{1}{2T_*}\sum_{R=R_s,R_o}\sqrt{1+\frac{4}{R-1}}\simeq0.93409438 ,
 \end{aligned}
 \label{eq:BR}
\end{equation}
for finite $N$, with the second argument held fixed. At given impact offset, the finite-spin clock with both finite endpoints accumulates fewer half-periods than $J_0/b$ by $B_R$ to leading order. Reaching the same exact order therefore requires a smaller impact offset, represented by the larger model argument $s+bB_R$. The sign-dependent phase offsets cancel at first order on summing both complete bins (Appendix~\ref{app:clock}). Table~\ref{tab:areas} compares this shift with the leading and fixed-spin predictions. At $\kappa=10^{-3}$ and $n=30$, the fixed-spin model is closer than the shifted prediction in the critical scene (errors $+0.61\%$ and $-3.59\%$, respectively).

\begin{table}[tbp]
 \caption{Prediction errors $100(G_{\rm pred}/G_{\rm ray}-1)$ in percent, alongside the fractional quadrature refinement difference $\epsilon_Q$. Critical rows use $\nu=0$. The fixed-spin logarithmic clock uses the spherical period, launch phases, and the same width cutoff.}
 \label{tab:areas}
 \begin{ruledtabular}
 \begin{tabular}{lccrrrr}
 Inclination & $\kappa$ & $n$ & Leading & Shifted & Fixed spin & $\epsilon_Q$\\
 \midrule
 Critical    & $10^{-3}$ & 20  & $+34.14$ & $-4.19$   & $+6.45$ & $4.9\times10^{-13}$\\
 $60^\circ$   & $10^{-3}$ & 20  & $+50.38$ & $-1.67$   & $+7.02$ & $6.8\times10^{-8}$\\
 Critical    & $10^{-7}$ & 431 & $+1.57$  & $-0.0100$ & $+7.37$ & $2.6\times10^{-7}$\\
 $60^\circ$   & $10^{-7}$ & 431 & $+2.01$  & $-0.00281$& $+16.41$ & $1.3\times10^{-7}$\\
 \end{tabular}
 \end{ruledtabular}
\end{table}

At fixed $s=2$ we take $\nu=-3,-1,0,1,3$ at $n=20,80,320$, together with $60^\circ$ controls. The critical area ratios $G/(b^6C)$ range over $0.651$--$0.745$, $0.885$--$0.922$, and $0.967$--$0.979$ respectively. Four further scenes use $b=0.025$, $s=1,4$ and $\nu=\pm1$. Across these 22 comparisons the largest refinement difference is $3.90\times10^{-6}$. The finite-spin area deficits exceed this resolution diagnostic.

\begin{figure}[tbp]
 \centering
 \includegraphics[width=\linewidth]{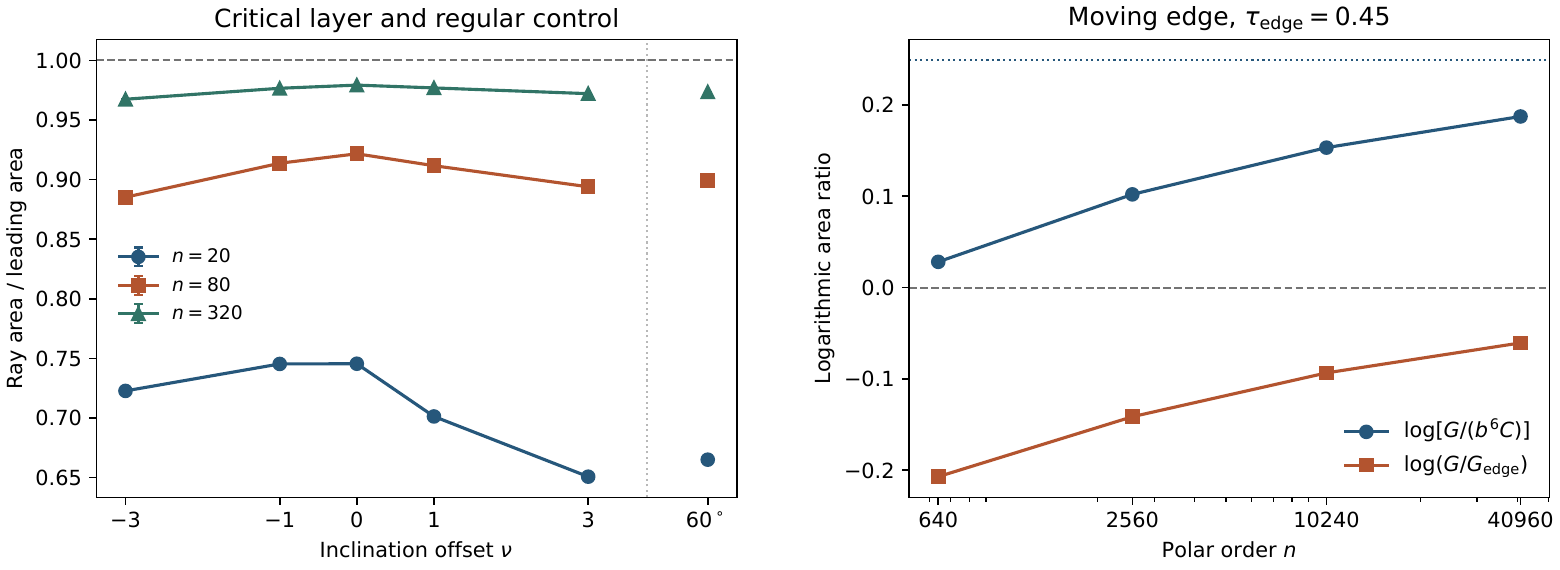}
 \caption{Finite-spin area checks at actual $s=2$. The left panel compares $G/(b^6C)$ across the signed inclination offsets for three orders. The isolated points use $G/(b^7C_{\rm ord})$ at $60^\circ$. Bars show the ray refinement differences and are smaller than the symbols. The right panel compares the critical and moving-edge predictions along $b^2L^{3/2}=0.45$, with the dotted line from Eq.~\eqref{eq:neg-overlap}.}
 \label{fig:crossover}
\end{figure}

Let $G_{\rm edge}$ denote the right side of Eq.~\eqref{eq:edge-area}. A comparison retaining finite factors is
\begin{equation}
 G_{\rm fe}=
 2(S_oB_o)_e A_\kappa(x_e)\sinh\gamma_e
  \sqrt{\frac{\pi\ell_e}{\Theta_x(x_e,0)}}\,e^{-n\gamma_e}.
 \label{eq:finite-edge-area}
\end{equation}
Here the factors are evaluated at the spherical edge and $A_\kappa$ is the exact-endpoint logarithmic intercept in Eq.~\eqref{eq:intercept}. The $2\sinh\gamma_e$ factor integrates both finite bins at the fold. Along $s=2$, $\tau_{\rm edge}=0.45$, Fig.~\ref{fig:crossover} and Table~\ref{tab:edge-checks} show a gradual rise of $\log[G/(b^6C)]$ towards the prediction $0.24882$, and a decreasing relative deficit against $G_{\rm edge}$. The log-width integration resolves widths below the floating-point range, with area refinement differences under $1.4\times10^{-12}$ on this sequence. The two-leg integration and outward budgets are given in Appendix~\ref{app:numerical-budget}.

\begin{table}[tbp]
 \caption{Moving-edge checks at $s=2$, $\tau_{\rm edge}=0.45$. The errors $\Delta_e$ and $\Delta_{\rm fe}$ are $100(G/G_{\rm pred}-1)$ in percent. The last three columns give the maximum CDF discrepancy on 551 thresholds in $[-0.5,5]$ for $(D_t-8)/4$ and $(D_t-I_e)/B_t$, and the largest CDF change on refinement across the two normalizations.}
 \label{tab:edge-checks}
 \begin{ruledtabular}
 \begin{tabular}{rrrrrrr}
 $n$ & $\log[G/(b^6C)]$ & $\Delta_e$ & $\Delta_{\rm fe}$ & $d_{\Gamma}$ & $d_{\rm sph}$ & $\epsilon_F$\\
 \midrule
 640   & 0.02820 & $-18.70$ & 1.628 & 0.343 & 0.115 & 0.0030\\
 2560  & 0.10210 & $-13.19$ & 0.645 & 0.274 & 0.098 & 0.0026\\
 10240 & 0.15329 & $-8.92$  & 0.257 & 0.222 & 0.081 & 0.0019\\
 40960 & 0.18740 & $-5.88$  & 0.102 & 0.179 & 0.066 & 0.0016\\
 \end{tabular}
 \end{ruledtabular}
\end{table}

The full-band geometric area is enclosed using outward-rounded ball arithmetic \cite{JohanssonArb}. At $\kappa=10^{-3}$, $n=20$, $\nu=0$, the interval $[1.5948455858,1.6095650373]\times10^{-8}\,{\rm sr}$ contains the independent ray value $1.60219513\times10^{-8}\,{\rm sr}$. For the equatorial cap of radius $0.4$ rad centered at longitude zero, the mixed-precision calculation gives an occupation interval of $[3.746353\%,4.031630\%]$, with relative half-width $(U-L)/(U+L)$ below $3.668\%$. The direct classifier uses binary64 with explicit margins in addition to Arb enclosures. Appendix~\ref{app:numerical-budget} gives the scope of the numerical guarantee. Independent ray quadrature gives $3.85155\%$.

At the moving edge, the full-band certificate in Table~\ref{tab:edge-certificate} covers three finite spins. For $\tau_{\rm edge}=0.45$ the relative half-width of the area is below $0.094\%$ at each spin. The ratios $G/G_{\rm fe}$ are also enclosed across the stated interval in $\tau_{\rm edge}$ at each spin. At $n=2560$ a finer fixed-parameter quadrature gives the tighter area bound $[2.76625,2.76796]\times10^{-199}\,{\rm sr}$, which contains the ray value $2.76710499\times10^{-199}\,{\rm sr}$.

\begin{table}[tbp]
 \caption{Full-band moving-edge area enclosures at $s=2$. The first two interval columns use $\tau_{\rm edge}=0.45$. The last encloses the ratio uniformly on $0.449\le\tau_{\rm edge}\le0.451$ at each stated spin.}
 \label{tab:edge-certificate}
 \begin{ruledtabular}
 \begin{tabular}{rrccc}
 $n$ & $b^{-1}$ & $\log(G/{\rm sr})$ & $G/G_{\rm fe}$ at $0.45$ & $G/G_{\rm fe}$ on the window\\
 \midrule
 2560  & 1280  & $[-457.19757,-457.19569]$   & $[1.00551,1.00740]$ & $[1.00323,1.00969]$\\
 10240 & 5120  & $[-1127.30082,-1127.29897]$ & $[1.00164,1.00349]$ & $[0.99961,1.00554]$\\
 40960 & 20480 & $[-2811.02545,-2811.02359]$ & $[1.00010,1.00195]$ & $[0.99816,1.00390]$\\
 \end{tabular}
 \end{ruledtabular}
\end{table}

With the limiting normalization $(D_t-8)/4$, the maximum discrepancy on the CDF grid occurs at the limiting onset $D_t=8$. At $n=2560$ about $0.274$ of the finer-resolution sample weight lies below that value. The finite spherical intercept is $I_e=7.68102$. Its shift from $8$ accounts for part of the weight below the limiting onset. Using both the finite intercept and the finite slope reduces the maximum discrepancy at this order to about $0.098$. The grid discrepancies fall along the sequence, reaching $0.179$ and $0.066$ at $n=40960$ for the limiting and finite spherical normalizations, respectively.

The strip integrations in Fig.~\ref{fig:coherence} cover $0\leq z_e\leq36$. At $n=2560$, restricting these integrations to $0.025^2\leq z_e\leq12$ would remove about $2.8\%$ of the uniform strip weight and $4.6$--$4.7\%$ with the shifted longitude window. At the same scene the outward area charge for $z_e>36$ is below $2.28\times10^{-16}G_{\rm fe}$. It bounds the omitted area and Fourier numerator for either profile. For a true retained strip area $A_{\rm in}$ the normalized change is bounded by twice this charge divided by $A_{\rm in}$. The other scenes use numerical positive tail integrals. Terminal phase residuals are below $3\times10^{-11}$ half-period units, and doubling the radial rule at selected rays changes the phase by at most $1.1\times10^{-4}$ rad. The spread between the plotted scrambles measures sampling sensitivity.

At $n=40960$ the two-scramble means for the localized strip are approximately $0.539-0.649\mathrm{i}$ and $-0.095-0.430\mathrm{i}$ for $m=1,2$, compared with limiting values $0.646-0.544\mathrm{i}$ and $0.075-0.436\mathrm{i}$. Their magnitudes are $0.844$ and $0.440$ compared with limits $0.845$ and $0.443$, and their phase offsets from the limits are $-0.178$ and $-0.387$ rad. Part of the phase drift comes from the radial gap near the turning point. For example, evaluating the two subtracted radial legs with the extremal spherical kernel at the finite $x_e$ gives a gap smaller than $J_*$ by $0.334$ at this order. The first harmonic for uniform longitude is below $0.002$ in both scrambles.

\begin{table}[tbp]
 \caption{Harmonics from fixed equatorial annuli with $W=1$ and the same longitude window as Fig.~\ref{fig:coherence}. The limiting weight and magnitudes follow from Eq.~\eqref{eq:disk-harmonics}. The finite entries are the rounded spans of the two scrambles at $n=40960$, $s=2$, $\tau_{\rm edge}=0.45$.}
 \label{tab:disk-harmonics}
 \begin{ruledtabular}
 \begin{tabular}{ccrrrrr}
 $[R_-,R_+]$ & $\mathcal T$ & $(A_{\rm d}/G)_\infty$ & $|c_1^{\rm d}|_\infty$ & $|c_1^{\rm d}|$ & $|c_2^{\rm d}|_\infty$ & $|c_2^{\rm d}|$\\
 \midrule
 $[9.5,10.5]$ & $1$   & $0.00081$ & $0.832$ & $0.832$ & $0.417$ & $0.416$--$0.417$\\
 $[3,18]$     & $1$   & $0.02440$ & $0.365$ & $0.365$--$0.370$ & $0.104$ & $0.111$\\
 $[1.24,10]$  & $1$   & $0.11801$ & $0.544$ & $0.528$--$0.534$ & $0.202$ & $0.183$--$0.200$\\
 $[1.24,10]$  & $1/2$ & $0.05338$ & $0.632$ & $0.620$--$0.627$ & $0.230$ & $0.214$--$0.231$\\
 $[1.24,10]$  & $0$   & $0$       & \multicolumn{4}{c}{Zero leading returning weight}\\
 \end{tabular}
 \end{ruledtabular}
\end{table}

The first two limiting inbound crossings are at $R\simeq1.95164$ and $1.18884$. The two outer annuli in Table~\ref{tab:disk-harmonics} transmit all the rays in the limiting layer. The transparent annulus $[3,18]$ has about 30 times the weight of $[9.5,10.5]$. Its wider phase range lowers the normalized first-harmonic magnitude from $0.832$ to $0.365$. The inner annulus contains the first inbound crossing and adds a return blocker for $R>2.86139$. At $\mathcal T=1/2$ the extra return passage downweights the outer part of this annulus, reducing the total response while increasing the normalized first-harmonic magnitude. At $\mathcal T=0$ the leading returning measure vanishes.

We evaluate the two-leg clock at each sampled radius, with $z_e\le36$, on both screen signs. At $n=40960$ each scramble uses $2^{18}$ points per sign in the central radial stratum and separate samples at the two ends. A positive numerical tail integration gives an omitted area scale below $4.3\times10^{-16}G_{\kappa,n}$ for the transparent models at this order. Independent quadratures between fixed radii locate the intervening crossings and agree with the integer counts. The terminal phase residual is below $3.0\times10^{-11}$ half-period units, and doubling the radial rule at selected rays changes the centered phase by at most $1.1\times10^{-4}$ rad. The finite magnitudes follow the predicted change with radial geometry. The complex phases still have visible finite-spin offsets. For example, the wide transparent annulus has first-harmonic phase about $1.69$ rad at $n=40960$, compared with the limiting $1.86$ rad in the common phase convention.

\section{Conclusions}

The turning excursion sets the width in impact space, and the observer's projection determines its contribution to the sky area. When $n\kappa^{1/3}$ stays finite, the cubic radial term sets that width. The square-root projection near $i_*$ changes the area power from $\kappa^{7/3}$ to $\kappa^2$. At a moving subcritical edge, the least unstable visible anchors dominate the exponentially small area, which is sensitive to the finite-spin instability.

The same radial selection ties arrival time to the screen. Rays at larger absolute screen height have larger spherical anchors, where less coordinate time accrues per polar traverse. The exponentially decaying single-order width and the fold Jacobian produce a radial density proportional to $z_e^{-1/2}e^{-z_e}$. The square-root map to the two signed screen branches cancels the fold factor, yielding a Gaussian height distribution and a quadratic advance in arrival time. Longitude mixing across a full order allows a fixed footprint to change the area while preserving the leading normalized time law under the occupation and boundary conditions given above.

The folded phase of a thin flash ties arrival time to source longitude through the residual azimuth in the corotating frame. A finite normalized harmonic can survive with vanishing absolute weight. A disk adds phases from different emission radii, so a wider annulus can increase the geometric weight while reducing the normalized modulation. Attenuation at disk crossings changes the contributing radii.

The inclination crossover supplies a geometric correction for models of high-order subrings on long interferometric baselines near $i_*$ \cite{JohnsonEtAl,GrallaLupsascaMarrone,PaugnatEtAl}. At the moving subcritical edge, Eq.~\eqref{eq:screen-time} gives the geometric arrival-time distribution for a simultaneous flash within a single polar order at a finite observer. Within the concentrating layer, rays at larger absolute screen height arrive earlier at leading order. This ordering provides a geometric input for models of photon-ring correlations and light echoes \cite{HadarAutocorrelations,WongEchoes}. The exponentially small solid angle suppresses the absolute geometric response even when normalized harmonics survive. Predicting interferometric visibilities or measurable echoes requires propagation to a distant observer, emission and velocity fields, frequency shifts and radiative transfer, and an instrument response. In particular, emission with a correlation time longer than the echo delay can erase a secondary peak in a light-curve autocorrelation \cite{CardenasGammieLupsascaEchoes}.

\begin{acknowledgments}
The author gratefully acknowledges OpenAI and its Codex tool, which was used to perform most of the calculations in this work.
\end{acknowledgments}

\section*{Data availability}

The source code and numerical data supporting this work are publicly available in the versioned GitHub archive \cite{PanKerrReproduction}. The \href{https://github.com/renqipan/kerr-photon-shell-research/archive/9533599df6006be1c4cb8b5e9f75586698636802.zip}{commit-pinned ZIP} contains the fixed inputs, software versions, SHA-256 file manifest, interval-coverage records, and complex coefficients and sampling diagnostics for Fig.~\ref{fig:coherence} and Table~\ref{tab:disk-harmonics}. The replay scripts recompute the mixed-precision direct cap integral and combine it with archived outer bounds. They also recompute the area enclosures in Table~\ref{tab:edge-certificate}.

\appendix
\section{Physical support and relative area tails}
\label{app:bounds}

The exact bin requires a long radial lapse. We first bound the impact offsets that can supply this lapse and then control their contribution to the sky measure. Write $r=r_c+y$ and $z_0=\kappa^2/x^2\leq(3-x)/4$. The exact polynomial is
\begin{align}
 \RR(r_c+y)&=y^4+4r_cy^3+c_2y^2+c_1y+c_0,\nonumber\\
 c_2&=4r_ch-w\sqrt{r_c}(x+4)-w^2,\qquad
 c_1=-2r_c^{3/2}(x+2)w-2xw^2,\nonumber\\
 c_0&=-4r_c^{3/2}x(1-z_0)w+
       \bigl[1+(x^3-7x^2)/4+\kappa^2\bigr]w^2 .
 \label{eq:quartic}
\end{align}
Compare with $4[y^3+hy^2-w(y+x)]$, whose positive root $p_0$ satisfies $p_0^2(p_0+h)=w(p_0+x)\equiv C_0$. In particular $p_0\leq\sqrt w$. For $0<\kappa\leq0.01$, $0<w\leq0.01$, substitution using $w/(p_0+x)=p_0 t_0+z_0t_0^2$, $t_0=p_0/(p_0+x)$, gives $\RR(r_c+p_0/3)/C_0<-1/32$ and $\RR(r_c+4p_0)>0$. For the lower trial radius, use $1-z_0\geq r_c/4$ and maximize the resulting convex quadratic in $t_0$ at $0$ and $1$. For the upper trial radius, at $y=\xi p_0$, $\xi\geq1$, a lower bound on $\RR/C_0$ is $r_c(3.29878\xi^2-8.01044\xi-15.392)$. The quartic derivative has one positive zero, so the outer displacement $\rho=r_t-r_c$ is simple and $p_0/3<\rho<4p_0$.

In $y=\rho+u^2$, the exact square-root denominator is $\sqrt{A_1+B_1u^2+4r_tu^4+u^6}$, with $A_1=\RR'(r_t)$ and $B_1=\RR''(r_t)/2$. The root brackets give $A_1\geq C_0/(128p_0)$ and $B_1\geq C_0/(1024p_0^2)$, using $2\rho B_1>A_1$. The first three coefficients bound those of the regularized comparison cubic from below by a factor $2^{-14}$. The band bounds $|\ell|<7.2$, $q<28$ imply $T\geq\pi/18$ by the elliptic formula. Extending both legs to infinity gives $J_0(X,w/b^4)\geq c_0' b\Lambda/T$ with $c_0'=\pi/(2304T_*)$. Thus for $s\geq s_0>0$ and small $b$, each selected width obeys
\begin{equation}
 w\leq b^4\delta_{\widetilde s}(X)\leq D_0 b^4,\qquad
 \widetilde s=c_0's_0/2 .
 \label{eq:necessary-width}
\end{equation}
The uniform bound $D_0<\infty$ follows from $A\geq4\delta$ in Eq.~\eqref{eq:clock}. The tube has unique impact labels because $\ell_x<0$ at fixed $w$ and, at fixed $\ell$, the allowed widths form an interval on which $(\partial_w q)_\ell=-B_o/\ell_x>0$.

On any region $h\geq h_0>0$ fixed, the spherical instability stays finite and bounded away from zero. Equation~\eqref{eq:necessary-width} and the model tails make the widths exponentially small in $1/b$. If all widths there are bounded by $W$, its impact-space area is $O(W)$ since $B_o\lesssim h/x$ and $\int h\,\dif x/x=O(1)$ on the spherical arc. At fixed $\ell$, $\Theta_i$ is affine with unit slope in $q$. Integrating $\Theta_i^{-1/2}$ over a visible length $v_\ell$ costs at most $2\sqrt{v_\ell}$. Cauchy--Schwarz then bounds the sky area by $O(\sqrt W)$, including any distant visibility edge.

For small $h_0$, in the near chart put $S=h+\rho\asymp h$, $q_1=\rho/S$, $t_1=\rho/(\rho+x)$, and $d_0=2+q_1-t_1$. On the width envelope in Eq.~\eqref{eq:necessary-width}, implicit differentiation of Eq.~\eqref{eq:quartic} gives
\begin{align}
 \frac{\rho_v}{\rho}&=d_0^{-1}+O(S),\quad
 \frac{A_1}{4r_c\rho S}=d_0+O(S),\quad
 \frac{B_1}{4r_c S}=1+2q_1+O(S),\nonumber\\
 \frac{(A_1)_v}{4r_c\rho S}&=\frac{2(1+2q_1)}{d_0}-t_1+O(S),
 \qquad \frac{(B_1)_v}{4r_c\rho}=\frac{3}{d_0}+O(S).
\end{align}
Both coefficient derivatives are positive, with the leading expression for the first exceeding $1/2$. The derivative of $4r_t$ is positive and the integration endpoints recede. Keeping the $A_1$ term on $\sqrt\rho\le u\le2\sqrt\rho$ gives $-\Lambda_v\gtrsim S^{-1/2}$ at fixed $X$. Also $\Lambda\lesssim S^{-1/2}(1+|\log q_1|)$. The period term is relatively small because $T_v=O(w)$ and $w(1+|\log q_1|)=O(S^2)$. At fixed $Z$, the anchor drift is $x_v=-\Theta_v/\Theta_x$ with $\Theta_x\gtrsim h/x$ and $\Theta_v=O(w)$. Its contribution to the slope is $O(S^{1/2})$ using $|x\partial_x(\Lambda/T)|\lesssim S^{-1/2}(1+|\log q_1|)$ from the differentiated quotient. The launch phase at fixed $Z$ has derivative $O(w)$. The ordinary chart at fixed $X$ has the same phase bound away from the fold.

\emph{Near-chart inverse.}
On a compact positive $X$ interval for ordinary inclinations, or a compact fixed-$Z$ interval on the critical near sheet, there is one root of Eq.~\eqref{eq:event} per sign and $\eta\in[0,1]$ for small $b$. On the admitted near widths, at the respective fixed label, $-\partial_v\mathcal E\gtrsim h^{-1/2}$. To see existence, choose two positive model clock levels strictly bracketing the interval of $s$ and take their inverse widths as trials. The scaled finite clock converges uniformly at these trials and the endpoint phases contribute $O(b)$. The trials bracket the bin on their connected graph, which exists at fixed $Z$ by $\Theta_x>0$. The slope gives uniqueness throughout the wider near domain satisfying Eq.~\eqref{eq:necessary-width}. The same positive bounds give the coarea envelope even where a chart is truncated. Any far sheet is covered by the fixed-$h_0$ bound.

The resulting envelopes for the area normalized by $b^7$ in $(X,\eta)$ and by $b^6$ in $(Z,\eta)$ are, respectively, $(H/X)\sqrt H\,\delta_{\widetilde s}(X)$ and $\sqrt H\,\delta_{\widetilde s}(X(Z))$, up to constant factors. The near visibility expansion
\begin{equation}
 \Theta_i=\Theta_*(i)+8x-4\kappa^2/x^2+O(h^2+w)
\end{equation}
excludes small $X$ at bounded $\nu$ in the near fold chart, and gives $Z^2\asymp X$ at large $X$. Equation~\eqref{eq:modeltails} makes both envelopes integrable. Outside $R^{-1}\le X\le R^2$ the near area, divided by $b^7$ or $b^6$ respectively, is bounded by a polynomial in $R$ times $e^{-cR}$. For the positive overlap set $d_i=\Theta_*(i)$ and $Q_i=d_i/b^2\to\infty$. Above $X=A_0/\sqrt{Q_i}$ with $A_0$ sufficiently large, $\Theta_i\geq d_i/2+4x$ and the fixed-$X$ launch derivative $O(w/\sqrt{d_i})$ is negligible. Below this cut the fold range is $Z=O(\sqrt{Q_i})$ and the small-$X$ tail is exponentially small in $\sqrt{Q_i}$ after normalization by $b^7/\sqrt{d_i}$. The far area divided by each of these normalizations still vanishes exponentially in $1/b$. This proves the positive overlap by dominated convergence on the remaining chart.

\section{Clock tails and the first endpoint correction}
\label{app:clock}

Factor the quartic in Eq.~\eqref{eq:clock} as $4(u^2+d_1)(u^2+d_2)$ with $0<d_1<d_2$. Then $d_1+d_2=H+3d$, $d_1d_2=3d^2+2Hd-\delta$, and $J_0=2K(1-d_1/d_2)/(T_*\sqrt{d_2})$, with elliptic parameter as argument. With $a_s=sT_*\sqrt H$, evaluation at $d/H=e^{-2a_s}$ and $e^{-a_s/2}$ brackets the inverse in either tail. There $d/X$ is exponentially small up to powers of $H$, and $d_1/d_2\sim2d/H$, $d_2\sim H$. The elliptic logarithm $K(1-\epsilon)=\log(4/\sqrt\epsilon)+O(\epsilon|\log\epsilon|)$ \cite{NISTHandbook} gives $d/H\sim8e^{-a_s}$ and $\partial_{\log d}J_0\sim-1/(T_*\sqrt H)$. This yields Eq.~\eqref{eq:modeltails}. Differentiated tails retain exponential decay up to algebraic factors.

For the large-$s$ model peak, put $\psi=2T_*\sqrt H$ and $\zeta=(L-X_*)\sqrt{s\psi''(X_*)/2}$. The quadratic expansion of $\psi$ in Eq.~\eqref{eq:abel} yields an inclination profile proportional to
\begin{equation}
 \mathcal F(\zeta)=\int_0^\infty t^{-1/2}e^{-(\zeta+t)^2}\,\dif t
\end{equation}
after removing $s^{-1/4}e^{-s\psi(X_*)}$. The exponential gap outside the neighborhood of $X_*$ and the vanishing tails of the rescaled profile confine the peak and half-height crossings to bounded $\zeta$. The equation $\mathcal F''+2\zeta\mathcal F'+\mathcal F=0$ makes each stationary point a strict maximum, hence the peak is unique. The two half-height points give the width quoted in Sec.~\ref{sec:inclination} through $\nu=-g(L)/c_i$.

For the endpoint shift, take $R^{-1}\le X\le R^2$ and positive trial widths with $J_0$ in a compact positive interval. The displacement $p=Y_t-X$ obeys $p^2(p+H)=\delta(p+X)$ and $p\le\sqrt{\delta}$. With $t=(r-r_c)/b^2$, the exact potential $b^{-6}\RR/4$ takes the form $U_b(t)-\delta V_b(t,v)$. Equation~\eqref{eq:quartic} gives
\begin{align}
 U_b(t)&=r_c t^2(t+H)+b^2t^4/4,\nonumber\\
 V_b(t,v)&=r_c^{3/2}\left[X(1-b^2/X^2)+(2+x)t/2\right]
       +b^2\sqrt{r_c}(x+4)t^2/4\nonumber\\
 &\quad -b^2\delta[1+(x^3-7x^2)/4+b^6]/4
       +b^6\delta(Xt/2+t^2/4).
 \label{eq:factored-comparison}
\end{align}
Consequently $U_b/[t^2(t+H)]$ and $V_b/(t+X)$ differ from one by $O(b^2P(R))$ on bounded $t>0$ for some polynomial $P$. The same bounds apply after the first two logarithmic derivatives in $t,\delta,X$. The factorizations use the common factors $t^2$ or $\delta$ before comparison. Since the log-$t$ slope of $t^2(t+H)/(t+X)$ lies in $[1,3]$, the exact simple outer displacement $p_b$ satisfies
\begin{equation}
 |\log(p_b/p)|+|\partial_v\log p_b-\partial_v\log p|=O(b^2P(R)).
 \label{eq:relative-turn}
\end{equation}
This follows by bracketing the logarithmic root equation and differentiating it with its slope bounded below. The regularized exact denominator squared is $Q_b=A_b+B_bu^2+4r_tu^4+b^2u^6$ after $t=p_b+u^2$. The model has $A=4(3p^2+2Hp-\delta)\geq4p(p+H)$ and $\delta\le p(p+H)$. Substitution of Eq.~\eqref{eq:relative-turn} into the derivatives of Eq.~\eqref{eq:factored-comparison} gives
\begin{equation}
 |A_b-A|+|\partial_v(A_b-A)|\lesssim b^2P(R)A .
 \label{eq:relative-constant}
\end{equation}
The analogous errors are relative to $B=4H+12p$ and to $4$ for the other two coefficients. At fixed $Z$, composing with the near visibility inverse costs $\log(X/X_0)=O(b^2P(R))$ and $\partial_v\log X=O(b^2\delta P(R))$. The derivative bounds above apply to this composition before integrating.

Let $Q=A+Bu^2+4u^4$. The difference between the inverse square roots of $Q_b$ and $Q+b^2u^6$, and its $v$ derivative, are bounded by $O(b^2P(R))(Q+b^2u^6)^{-1/2}$. The latter has a bounded integral on the trial levels. For the remaining sextic-minus-quartic difference, the value after subtracting its pure quartic counterpart is bounded by $O(b^2)\min\{1,(A+Bu^2)/u^4\}$. Its $v$ derivative is bounded by $O(b^2)u^2/\sqrt Q$ for $u\le1$ and $O(b^2P(R))(1+u^2)/u^4$ for $u\ge1$. Both bounds are integrable since $Q\ge4u^4$. The pure-quartic calculation gives
\begin{equation}
 2\int_0^{U_R}\left[\frac{1}{\sqrt{4u^4+b^2u^6}}-\frac{1}{2u^2}\right]\dif u
 -2\int_{U_R}^\infty\frac{\dif u}{2u^2}
 =-\frac{\sqrt{4+b^2U_R^2}}{2U_R},\qquad U_R=\frac{\sqrt{R_j-r_t}}b .
\end{equation}
Summing the two legs gives $b\Lambda/T=J_0-bB_R+O(b^2P(R))$. The term $-bB_R$ is the net loss of scaled polar phase from the large-radius part of the clock. Motion of $U_R$ costs $O(b^3P(R))$ in the derivative since $\partial_v U_R=O(bP(R))$ and the integrand there is $O(b^2)$. The period and phase contributions are $O(b^2P(R))$ in the scaled slope, so $b\partial_v\mathcal E=\partial_v J_0+O(b^2P(R))$ in each fixed-label chart. At the exact root, the model level shifts by $b(B_R+j_\sigma-1/2+\eta-\varphi_\sigma^0)$ with limiting launch phase $\varphi_\sigma^0$. The sign sum and $\eta$ average give the $B_R$ shift since $j_++j_-=\varphi_+^0+\varphi_-^0=1$. Taking $R$ proportional to $1+|\log b|$ leaves $O(b^2)$ of the normalized area outside the common core. Coarea then gives Eq.~\eqref{eq:BR}.

\section{Relative estimates at the subcritical edge}
\label{app:edge}

\emph{Moving-edge chart.}
In Eq.~\eqref{eq:moving-limit}, visibility and Eq.~\eqref{eq:necessary-width} exclude $x<u_e/2$. On $u_e/2\le x\le r_0$ for fixed small $r_0$, $\kappa^2/x^3=O(L^{-3})$, $\Theta_x\gtrsim1$, and $w/x^2\to0$. The fixed-fold equation $\Theta_i(x,w)=\Theta_i(x_0,0)$ has a unique near solution with $|x-x_0|=O(w)$ and $x_v=O(w)$. The physical chart starts at $x_0=x_e$, with its zero-solid-angle endpoint interpreted by continuity. Let
\begin{align}
 d_c&=\frac{\sqrt{4r_c^{3/2}x(1-z_0)}}{\lambda_r},
 & F_c(y)&=\frac{2\lambda_r^2 y}
 {\lambda_r\sqrt{\lambda_r^2+4r_cy+y^2}+\lambda_r^2+2r_cy},\\
 A_\kappa(x)&=\frac{4F_c(R_s-r_c)F_c(R_o-r_c)}{d_c^2}.
 \label{eq:intercept}
\end{align}
The factor $A_\kappa$ records the regular part of the radial lapse after subtracting the divergence near the double root. With $p_t=(r_t-r_c)/x=d_c\sqrt w/x+O(w/x^2)$, split each radial leg at $y=x\sqrt{p_t}$. In the inner piece the scaled potential is quadratic to relative order $O(\sqrt{p_t})$ after the simple root is divided out. It contributes $\log(2/\sqrt{p_t})+o(1)$ to $\lambda_r\Lambda$ per leg. In the outer piece, replacing the potential by its spherical value costs $o(1)$, by the relative bound $O((w/x^2)/\xi^2+(w/x^2)/\xi+w/x)$ with $\xi=y/x$. Its integral is $\log[F_c(R_j-r_c)/F_c(x\sqrt{p_t})]$, with $F_c(x\sqrt{p_t})\sim x\sqrt{p_t}$. Consequently
\begin{equation}
 \lambda_r\Lambda=\log(A_\kappa(x)/w)+o(1),\qquad
 \lambda_r\Lambda_v=-1+o(1),\qquad
 |\lambda_r x\Lambda_x|\lesssim 1+|\log(w/x^2)| .
 \label{eq:edge-clock}
\end{equation}
For the width derivative the inner constant coefficient of the root quotient has logarithmic slope $1/2+o(1)$. Rescaling its regularized coordinate by $\sqrt{p_t}$ gives the integrable weight $(1+u^2)^{-3/2}$ and a contribution $-1/2$ per leg. The other differentiated terms are $O(p_t(1+|\log p_t|))$. The fixed-fold drift costs $O((w/x)(1+|\log(w/x^2)|))$ in the normalized clock and derivative. With $T-T_c=O(w)$ this gives $\gamma_\kappa(x_0)\partial_v\mathcal E|_{\Theta_i}=-1+o(1)$. For sufficiently small $r_0$, trial levels $\log(A_\kappa(x_0)/w)=n\gamma_\kappa(x_0)\pm1$ bracket every phase of each bin. The trial widths lie in the same small-$w/x^2$ chart and the normalized slope gives a unique inverse there. Since $A_\kappa(x_0)\sim64x_0^2$ as $x_0\to0$ and the bounded phase offsets cost $O(\gamma_\kappa(x_0))$ in the exponent, Eq.~\eqref{eq:edge-width-law} follows.

In the $(x_0,\eta)$ chart the one-sign density is
\begin{equation}
 \frac{S_oB_o}{\Theta_x(x,w)}
 \frac{\Theta_x(x_0,0)}{\sqrt{\Theta_i(x_0,0)}}\,
 \frac{w}{|\partial_v\mathcal E|_{\Theta_i}} .
\end{equation}
Use $w\lesssim x_0^2e^{-n\gamma_\kappa(x_0)}$, $\gamma_\kappa'(x)\gtrsim x^{-1/2}$, and $\Theta_i(x_0,0)\gtrsim x_0-x_e$ in the near region. In $t_e=n(\sqrt{x_0}-\sqrt{x_e})$ the density in $\dif t_e\,\dif\eta$, divided by $x_e^{11/4}n^{-1/2}e^{-n\gamma_e}$, is bounded by $C(1+t_e)^6t_e^{-1/2}e^{-c t_e}$. The far area $O(e^{-c_0/b})$ is negligible relative to this scale since $\gamma_e\to0$ and $x_e\gtrsim b^2$. These bounds justify the endpoint Laplace integral and bounded tests of the conditional law.

For Eq.~\eqref{eq:edge-expanded}, substituting $x=u_e$ makes the linear visibility terms cancel by $c_i(i-i_*)=-8u_e+4\kappa^2/u_e^2$. The coefficient of the residual $u_e^2$ is $6+4\cot^2 i_*+32\Theta_*''(i_*)/c_i^2$. Its negative divided by the limiting edge slope $8$ gives $K_e$. The angular period satisfies $\partial_q\log T|_{(1,2,3)}=-1/12$, from its elliptic integral, and $q_c=3+8x+O(x^2+xL^{-3})$ near the edge. Inserting this period correction into $2\sqrt{r_ch}\,T_c$ gives the expanded exponent.

At fixed subcritical $i$, the extremal visible interval is bounded away from $x=0,3$. Its endpoints are simple since the derivative of the extremal $\Theta_i(x,0)$ has one positive zero, while $\Theta_i$ is positive at $x=\sqrt 2$. The uniform logarithmic saddle law on a compact neighborhood gives $w=e^{-n\gamma_\kappa(x)+O(1)}$ for the exact bin. Integrating the visibility factors across both transverse edges gives the upper exponential rate. An interval in the visible interior with instability arbitrarily close to the minimum gives the lower rate.

\section{Averaging and the timing estimates}
\label{app:timing}

On a critical compact fold core the log-width derivative of $\kappa\phd$ tends to $\partial_{\log\delta}\Psi_0$ uniformly, by the same quotient comparison as for the clock with the additional integrable factor $Y^{-1}$. Dividing by the exact clock slope gives
\begin{equation}
 b^2\partial_\eta\phd\longrightarrow
 \left.\frac{\partial_{\log\delta}\Psi_0}{\partial_{\log\delta} J_0}
 \right|_{(X_0,\delta_s(X_0))}>0 .
\end{equation}
Both derivatives are negative because the turning coordinate and the regularized coefficients increase with width. To average a longitude test of zero mean, change variables from $\eta$ to longitude on any fixed subinterval. Replacing the inverse derivative by its constant leading value costs $o(1)$, and complete longitude periods cancel. The two incomplete pieces cost $O(b^2)$. Approximation of the polar test by step values on a fine $\eta$ partition then gives the source average along the limiting half-traverse. The leading lapse test is constant in $\eta$ on each fold label. Relative area tails and continuous upper and lower approximations to boundary-null masks finish the proof of Eq.~\eqref{eq:critical-product}.

For the moving-edge deficits, angular endpoint and period errors are $O(1+nw)$ before rescaling. The radial impact correction to $K_j$ is $O(w/x^2)$, with integrated scaled error $O(nw/\sqrt x)$. The subtracted spherical kernel multiplies $\Lambda=nT_c+O(1+nw)$, so replacing this lapse by $nT_c$ there costs $O(\sqrt{x}(1+nw))$ after scaling. The remaining numerator times $x$ is bounded by $C\xi/(1+\xi)$ in $\xi=(r-r_c)/x$. On $\xi\geq2p_t$ the scaled radial square root is bounded below by $c\xi\sqrt{1+\xi}$, and between $p_t$ and $2p_t$ by $c\sqrt{p_t(\xi-p_t)}$. The inner piece is $O(p_t)$ and the outer piece is dominated by $(1+\xi)^{-3/2}$. Its limit gives the stated $-k_j$ per leg. The center constants are $d_t=(16/3-4K_e)T_*+I_t$ and $d_\phi=(2/3-2K_e)T_*+I_\phi$, where $I_j=\int_{\rm half} A_j^*\dif\lambda$. They follow by expanding $K_t^c(r_c)=4/x+8+O(x+\kappa^2/x)$ and $K_\phi^c(r_c)=2/x+2+O(x+\kappa^2/x)$, together with the period and edge expansions.

For the within-bin slopes, set $r=r_c+x(p_t+u^2)$. The scaled positive root quotient has constant coefficient of order $p_t$ and all other coefficients bounded. Its width derivative localizes at $u=O(\sqrt{p_t})$, where $xK_j\to k_j$. The other terms in the derivative of the normalized lapse are $O((p_t+w/x)(1+|\log p_t|))$, including fixed-fold drift. Thus $\lambda_r x\partial_v\Mtime\to-4$ and $\lambda_r x\partial_v\phd\to-2$. Dividing by the clock slope proves Eq.~\eqref{eq:bin-speed}. The within-bin longitude average uses the same partition argument, now on the scale $x_e^{-1}$.

At fixed source latitude in the thin strip and on $z_e$ in a compact subinterval of $(0,\infty)$, \mbox{Eq.~\eqref{eq:edge-clock}} and the smooth terminal polar phase give $\ell_e\partial_{x_0}v=O(1)$ and $\partial_{x_0}x=1+o(1)$. For the two-leg remainder from subtracting the spherical kernel, write the lapse as $nf_j(x)+x^{-3/2}\mathcal J_j$ plus the angular endpoint and impact corrections. The normalized subtracted numerator in the regularized quotient is bounded by $C[\min(1,p_t+u^2)+w/x]$, also after a fixed-width log-radius derivative. Hence $\mathcal J_j$ and its log-radius derivative are bounded, and its log-width derivative is $o(1)$ by localization at $u=O(\sqrt{p_t})$. The scaled derivative in $z_e$ of the remainder tends to zero, since $\ell_e/x\to0$, and the scaled angular and impact derivatives vanish there. The leading derivative is that of $nf_j(x)$, giving $x_e^{3/2}\partial_{z_e}\Mtime\to-4$ and $x_e^{3/2}\partial_{z_e}\phd\to-2$. This gives radial longitude averaging at each strip latitude.

For the gap kernel $K_D=K_t-2K_\phi$, the spherical derivative in radius is uniformly bounded when $\kappa^2/x^3\to0$. The impact displacement costs $O(nw/x)=o(1)$ in the integrated gap. After subtracting its spherical value at $r_c$, the radial integrand is bounded by $C(r-r_c)^{-1/2}$ outside the adjacent simple-root interval, which contributes $O(\sqrt{x}\,p_t)$. At fixed $r>1$ the kernel tends to $(r+1)^2$, its spherical value tends to $4$, and the potential tends to $(r-1)^3(r+3)$. The two-leg integral therefore tends to $J_*$. The angular gap kernel is even about the equator, so a turn to an equator crossing contributes half of a half-period integral. Combining the angular and radial contributions gives Eq.~\eqref{eq:two-gap}, since $H_\kappa'(x)=O(1)$ and $n\ell_e\to0$. The null first variation at fixed endpoint radii and latitudes gives $\dif\Mtime=\ell\,\dif\phd$ \cite{KovnerFermat}. It also gives the vanishing gap shear, $\partial_{z_e}D=(\ell-2)\partial_{z_e}\phd\sim2\sqrt{x_e}$, since $\ell-2\sim-x_e^2$ on the radial core.

The phase-to-latitude change near the equator contributes $1/(T_*\sqrt3)$ and restricts each sign to $O(x_e)$ of the bin. This controls the omitted near-strip mass by $O(x_e)$ times the radial tails of Appendix~\ref{app:edge}. The far bound is $o(x_eG_{\kappa,n})$. The radial longitude average, the equal limiting sign weights, and the phase shifts $\pm H_*/4$ give Eq.~\eqref{eq:harmonics} with any bounded continuous test of $z_e$, proving independence.

\section{Disk area, blocking, and phase}
\label{app:disk}

Fix the spherical fold label $x_0$ and the observer sign. Equation~\eqref{eq:disk-phases} gives $\partial_R \mathcal E_{\rm ret}=1/(T\sqrt{\RR(R)})$ at fixed width. The exact crossing fixes $\mathcal E_{\rm ret}=n+j_\sigma$, so increasing $R$ requires a wider impact offset to compensate for the added radial lapse,
\[
 \left.\frac{\dif w}{\dif R}\right|_{x_0}
 =-\frac{w}{(\partial_v \mathcal E_{\rm ret})_{x_0,R}}\,\frac{1}{T\sqrt{\RR(R)}}>0
\]
in the moving-edge chart. Equations~\eqref{eq:intercept} and \eqref{eq:edge-clock} hold with $R$ in place of $R_s$, uniformly on the fixed closed annulus. On compact $z_e$ intervals the spherical intercept is asymptotic to $64x_0^2$ uniformly in $R$. The clock derivative has leading value $-1/\gamma_\kappa(x_0)$. The sky measure thus has the same $z_e$ density and asymptotically equal sign weights as the sphere, multiplied by $\dif R/(T_*\sqrt{\RR_*(R)})$. The term $\dif R/\sqrt{\RR(R)}$ is bounded on the chosen annulus for sufficiently small $x$. The near tail is dominated by the integrable bound in Appendix~\ref{app:edge} times this bounded annular measure. To bound the far tail, apply the necessary lapse inequality with the largest permitted endpoint $R_+$, which only increases the clock. The positive-quotient width bound and the full far sky estimate then give exponentially small area relative to $G_{\kappa,n}$. An inbound ray outside the concentrating layer, if present, also satisfies this necessary inequality, since the return to $R_+$ adds a nonnegative lapse.

For any finite radii in the annulus, the inbound clock tends to $\mathcal Y(R)-\mathcal Y(R_o)$, since $\mathcal Y'(R)=-1/\sqrt{\RR_*(R)}$. On the return, the phase elapsed between $R_-$ and the terminal integer at $R$ tends to $[\mathcal Y(R_-)-\mathcal Y(R)]/T_*$. These intervals give Eq.~\eqref{eq:disk-blockers}. The outgoing blocking transitions occur at isolated radii. Their neighborhoods carry arbitrarily small radial measure, so they do not affect the limiting integral for bounded $W$.

The gap estimate in Appendix~\ref{app:timing} holds uniformly on the annulus, with the source-leg residual given by $J(R)$ instead of $J(10)$. At fixed $R$ the radial derivative of the reception azimuth satisfies $x_e^{3/2}\partial_{z_e}\Psi\to-2(1-2\Omega_*)$, by the same subtracted-kernel estimates with a fixed terminal equator phase. On compact radial-layer intervals the inverse derivative is asymptotically constant at its leading scale. Complete longitude cycles therefore average $f(\phi_s)e^{-im\phi_s}$ and $f(\phi_s)$ separately, and the incomplete cycles have vanishing measure. Removing the interval cutoffs with the relative tail bound and averaging the two signs gives Eq.~\eqref{eq:disk-harmonics}.

\section{Logarithmic quadrature and finite interval budgets}
\label{app:numerical-budget}

\emph{Moving-edge ray integration.}
Set $v=\log w$, compute the exact outer root as $r_t-r_c=e^{v/2}p_v$, and solve Eq.~\eqref{eq:quartic} divided by $w$ for $p_v$. For $A_1=\RR'(r_t)$ and $B_1=\RR''(r_t)/2$, keep $\log A_1=v/2+\log(A_1e^{-v/2})$. The root quotient of Appendix~\ref{app:bounds} admits the coordinate $\xi=\operatorname{asinh}(\sqrt{B_1/A_1}\sqrt{r-r_t})$. Writing $u^2=(A_1/B_1)\sinh^2\xi$, computed from logarithms, gives
\begin{equation}
 \frac{\dif r}{\sqrt{\RR(r)}}=
 \frac{2\,\dif\xi}{\sqrt{B_1+u^2(4r_t+u^2)\tanh^2\xi}} .
\end{equation}
The regularized lapse retains the two exact upper endpoints. Finite-width changes to the angular factors in the fixed-fold chart are below working precision in the four deep-edge scenes. The polar clock is inverted with its computed residual. A bin with $v_1<v_0$ contributes width $e^{v_0}[-\operatorname{expm1}(v_1-v_0)]$. Extracting the common scale $A_\kappa(x_e)e^{-n\gamma_e}$ keeps relative width weights in range even when the area itself is stored only as a logarithm. In the critical coefficient the deep-tail calculation similarly extracts $e^{-2sT_*\sqrt L}$. The 55-digit spot checks use a separate scaled root, two lapse integrals, and angular primitives. At $n=40960$ a checked ray has $w\simeq10^{-1214.91}$ and $r_t-r_c\simeq10^{-607.46}$.

\emph{Critical benchmark.}
At $\kappa=10^{-3}$, $n=20$, $\nu=0$, the $Z$ chart from 0 to 24 has 16,384 intervals per sign with strictly positive lower bounds on $-\partial_v\mathcal E$ at fixed $Z$ throughout the verified width intervals. The accumulated chart bounds are approximately $1.59484558583\times10^{-8}$ and $1.60956500729\times10^{-8}$ sr. The complement of the width rectangles on this chart is excluded by 98,304 strict phase bounds, with $T$ and launch phase included in the endpoint tests. Outside the chart, 256 radial boxes use the necessary condition $\Lambda/T\ge n-1$ and positive area charges for a total below $2.997\times10^{-16}$ sr. The full-band value includes this upper charge. For the cap we use a second exact chart on $0\le Z\le8$. Let $p=r_t-r_c$, so that the turning equation has the form $a_0-a_1w+a_2w^2=0$ at fixed $(x,p)$. The physical root is $2a_0/(a_1+\sqrt{a_1^2-4a_0a_2})$. At fixed $(Z,\log p)$ we then solve the fold equation for $x$, and use its derivative and the radial turning equation for the sky Jacobian. The new rectangles cover the projections of the certified width intervals for both signs in this label range. Positive derivatives of the fold equation identify their roots throughout the covering rectangles.

On each patch, a complex contraction bounds the analytic continuation of the fold root on the product of two Bernstein ellipses of parameter 4. The clock and azimuthal lapse are bounded there using the positive outer-root quotient with fixed integration coordinates. The normalized clock, azimuth without its terminal polar primitive, and sky density are interpolated on 15 Chebyshev nodes in each coordinate, with every node evaluated in ball arithmetic. A function bounded by $M_f$ on the product ellipses has tensor interpolation error at most $16M_f\,4^{-15}/(1-1/4)^2$. We enclose the terminal polar contributions separately in Arb before exporting the angle tables. The direct classifier then evaluates the polynomials in binary64, selects lookup ranges, reduces longitude modulo $2\pi$, clips against an affine phase on unresolved cells, and accumulates positive contributions. The interpolation error includes the sum of the coefficient conversion errors. The implementation also adds $10^{-10}(1+\sum_{ij}|c_{ij}|)$ to that error, applies a $10^{-8}$ rad phase pad in clipping, and expands the patch sums by relative factors $1\pm10^{-8}$ before the final Arb assembly. Taylor bounds and angular allowances account for the departure from an affine phase and for the latitude dependence of the rim. The \href{https://github.com/renqipan/kerr-photon-shell-research/blob/9533599df6006be1c4cb8b5e9f75586698636802/reproducibility/methods.md\#critical-cap}{archived methods} list the numerical margins. An end-to-end rounding-error bound for the binary64 classifier is unavailable. The 808 patches give the mixed-precision interval $[6.0299992,6.3490714]\times10^{-10}$ sr on the direct part. On $8\le Z\le24$ the azimuth average and variation give an upper charge below $8.076\times10^{-12}$ sr. The cap area outside the main chart is bounded by $2.518\times10^{-17}$ sr. Dividing by the full-band geometric interval gives the fraction in Sec.~\ref{sec:numerics}.

\emph{Moving-edge certificates.}
For the three spins in Table~\ref{tab:edge-certificate}, visibility excludes $x$ below $0.004$, $0.0016$, and $0.00065$, respectively. From those floors to $x=0.1$, the radial quotient and lower period bound give $w\le 3925h^2 e^{-1.13381n\sqrt h}$, with $\kappa^2/x^3<64^{-6}$ and $n\sqrt h>128$. All three tubes satisfy $w/x^2<10^{-72}$, $\Theta_x>7$, and $|\Theta_w|<10$, so the fixed-fold drift is bounded by $2w$. Splitting the scaled radial lapse at $(r-r_c)/x=10^{-9}$ bounds the clock discrepancy relative to the logarithmic intercept by $1.21\times10^{-8}$ in $\lambda_r\Lambda$. We allow $10^{-6}$ in log width for the clock and $10^{-9}$ for the period, phase and label corrections. The positive central bin width is multiplied outward by $(e^{\pm\epsilon_v}-r e^{\mp\epsilon_v})/(1-r)$ with $\epsilon_v=10^{-6}+10^{-9}$ and $r=e^{-\gamma_{\min}}$, using the lower instability bound on the near chart.

Positive ball integrals factor out $e^{-n\gamma_e}$ in the variable $u=\sqrt{z_e}$. The core $0\le u\le5$ supplies the lower and upper contributions. The remaining near cells give a positive upper charge and extend past $x=0.09$. Far cells from $x=0.09$ to $3$ use the integrated positive quadratic bound on the root quotient and the far solid-angle bound. At each spin the window is covered by 32 parameter cells. Within a cell, evaluating the instability difference as the integral of $\gamma_\kappa'$ from the uncertain edge preserves the common edge in the exponent. For the spherical strip at $n=2560$, a separate positive integral starting at the inward-rounded width-chart cut $u=191/32$ gives the tail bound at $z_e=36$ in Sec.~\ref{sec:numerics}.

For the disk comparisons, the source endpoint in the log-width quadrature is replaced by the sampled $R$. The radial weight uses $\dif w/\dif R$ from Appendix~\ref{app:disk} with its logarithmic inverse slope. At fixed samples, a centered log-width difference checks this slope to relative accuracy better than $5.5\times10^{-11}$ at $n=40960$. The occultation count uses a separate radial quadrature through the annulus, and bisection in the physical radius checks individual crossings.

\bibliographystyle{apsrev4-2}
% Allow a short line before an arXiv identifier without stretching the entry.
\newcommand{\Eprint}[2]{\hfil\penalty0\hfilneg\href{#1}{#2}}
\bibliography{references}

\end{document}